\documentclass[%
 aip,
jap,
 amsmath,amssymb,
 reprint,%
]{revtex4-2}

\usepackage{graphicx}
\usepackage{dcolumn}
\usepackage{bm}
\usepackage{physics}
\usepackage{setspace}
\usepackage{float}
\usepackage[utf8]{inputenc}
\usepackage{mathptmx}
\usepackage{hyperref}
\usepackage{xcolor}
\newcounter{saveenum}

\begin{document}

\preprint{AIP/123-QED}

\title{DECaNT: Simulation Tool for Diffusion of Excitons in Carbon Nanotube Films}
\author{S. W. Belling}\altaffiliation{SWB, YCL, and AHD contributed equally to this work}\email{swbelling@wisc.edu}
  \affiliation{Department of Electrical and Computer Engineering, University of Wisconsin--Madison, Madison, Wisconsin 53706}
\author{Y. C. Li}\email{li779@wisc.edu}
  \affiliation{Department of Electrical and Computer Engineering, University of Wisconsin--Madison, Madison, Wisconsin 53706}
 \author{A. H. Davoody}
 \altaffiliation{Presently with Intel Corporation, Hillsboro, Oregon}
 \affiliation{Department of Electrical and Computer Engineering, University of Wisconsin--Madison, Madison, Wisconsin 53706}
  \author{A. J. Gabourie}
  \altaffiliation{Presently with Stanford University, Stanford, California}
 \affiliation{Department of Electrical and Computer Engineering, University of Wisconsin--Madison, Madison, Wisconsin 53706}
  \author{I. Knezevic}\email{iknezevic@wisc.edu}
 \affiliation{Department of Electrical and Computer Engineering, University of Wisconsin--Madison, Madison, Wisconsin 53706}

\date{\today}
\begin{abstract}
We present the numerical tool DECaNT (Diffusion of Excitons in Carbon NanoTubes) that simulates exciton transport in thin films of carbon nanotubes. Through a mesh of nanotubes generated using the Bullet Physics C++ library, excitons move according to an ensemble Monte Carlo algorithm, with the scattering rates that account for tube chirality, orientation, and distance. We calculate the diffusion tensor from the position--position correlation functions and analyze its anisotropy and dependence on the film composition, morphology, and defect density.
\end{abstract}

\maketitle

\section{Introduction}
Carbon nanotube (CNT) films are promising materials for a number of applications.~\cite{Jariwala_2013_cnt_applications_rev,Arnold_2013_cnt_review,Laird_review_2015} Their use in energy-harvesting photovoltaics~\cite{Arnold_Films_Harvesting_2011,Jeon_cnt_transparent_electrode_2015,near_IR_photovoltaics} hinges on understanding the transport of photogenerated excitons (bound electron--hole pairs) both within a single tube and among different tubes in an ensemble. Experiments have shed light on exciton lifetime, diffusion length, and quantum efficiency in semiconducting CNT films.~\cite{flach_2020_improved_exciton_liftime,Ruzicka_2012_exciton_diffusion_constant,Shea_2013_exciton_diff_length} Measurements of exciton lifetime and diffusion length are particularly useful for quantifying CNT-film performance in photovoltaics, because, in typical designs, excitons must traverse the film and arrive at the $\mathrm{C}_{60}$ layer for harvesting.~\cite{CNT_c60_interface_Dowgiallo_2016,Shea_2013_exciton_diff_length} Typical exciton diffusion lengths are of order nanometer and lifetimes of order picosecond.

Microscopic theory and simulations benchmarked against experiment can be invaluable in deepening the understanding of nanoscale systems, such as CNT films.~\cite{Postupna_2014_JPCL,Wong_2009_JCP,Jones_2019_RET_review} Previously, we theoretically studied the transfer of excitons between two-tube CNT systems of varying separation, orientation, and chirality.~\cite{Davoody_exciton_transfer,Davoody_exciton_phonon_transfer} We found that environmental disorder has a large impact on transfer between tubes of the same chirality, with tubes showing different behaviour depending on whether the CNTs are parallel to one another or not.~\cite{Davoody_exciton_transfer} We also studied the effect of phonon scattering on exciton transfer and showed that inelastic scattering mediated by phonons enables nonresonance energy transfer (transfer between exciton states of different energies in the donor and acceptor tubes), as well as transfer between bright and dark states.~\cite{Davoody_exciton_phonon_transfer}
But, even with detailed knowledge of exciton dynamics in binary CNT systems, their transport through CNT films is challenging to simulate because it depends on film composition, morphology, and even processing steps that may affect the dielectric environment and introduce defects.

In this paper, we present the numerical tool DECaNT (Diffusion of Excitons in Carbon NanoTubes)~\cite{DECaNT} that simulates exciton transport in CNT films. The simulation accounts for the factors relevant in experiment, such as film composition, morphology, and defect density. We utilize the Bullet Physics C++ library ~\cite{bullet_physics} to generate a three-dimensional (3D) mesh of CNT segments with extraordinary control over the film morphology. In DECaNT, the dynamics of excitons scattering between tubes is simulated using the ensemble Monte Carlo (EMC) technique, which incorporates our previously published bandstructure and transfer-rate calculations.~\cite{Davoody_exciton_transfer,Davoody_exciton_phonon_transfer} We track the position of each exciton in the film and calculate the full diffusion tensor from the position--position correlation functions. We analyze the impact of tube chirality, intertube spacing, orientation, and defect density on the diffusion coefficient and length.

The paper is organized as follows. In Sec.\ref{sec:exciton_physics}, we review the physics of exciton transfer in CNTs. In Sec. \ref{sec:mesh_generation}, we discuss the generation of the CNT film and different film morphologies. In Sec. \ref{sec:EMC}, we explain the implementation of the Monte Carlo technique for calculating the diffusion tensor that quantifies exciton motion through the aggregate. In Sec. \ref{sec:Results}, we show calculations of the diffusion tensor for representative film morphologies and study the effects of intertube spacing and quenching-site density on diffusion properties. We conclude with Sec. \ref{sec:conclusion}.

\section{Exciton Physics in Carbon Nanotubes}\label{sec:exciton_physics}

\begin{figure}
\begin{center}
\includegraphics[width=\linewidth]{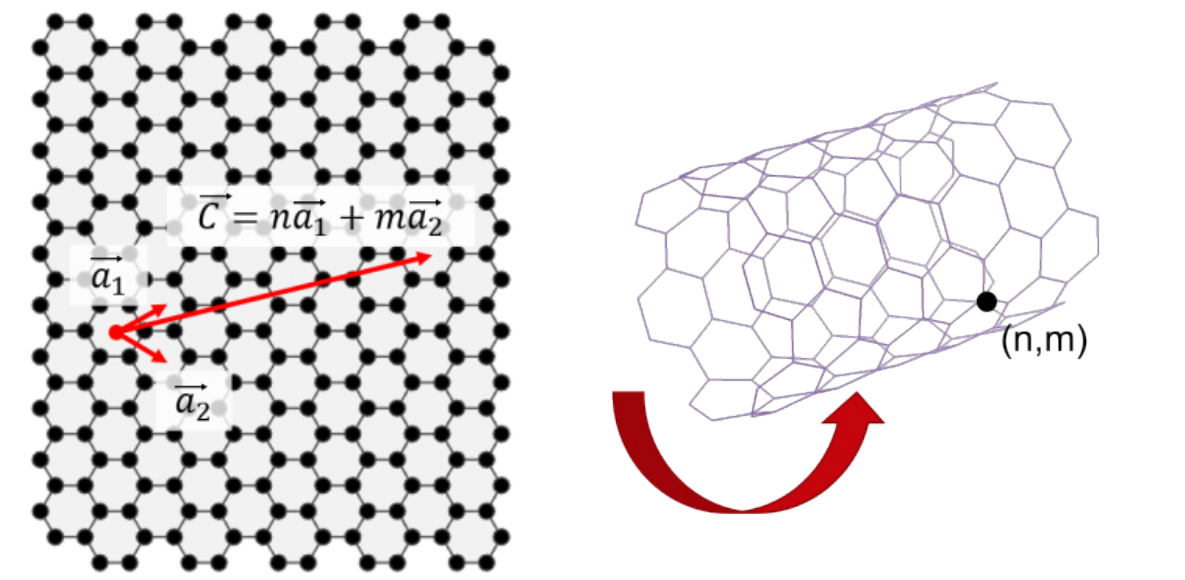}
\caption{A graphene sheet being rolled into a CNT. The chiral vector $\vec{C}=n\vec{a}_1+m\vec{a}_2$ is a linear combination of lattice vectors and denotes the rolling direction. The ordered pair of integers $(n,m)$ is referred to as the tube chirality. CNTs with different chiralities have different bandstructures and electronic properties.}\label{Fig1}
\end{center}
\end{figure}

Optical excitation of the electronic system in CNTs results in the formation of Frenkel excitons, tightly bound electron--hole pairs that can travel nanometer distances before dissociating. ~\cite{Avouris_CNT_Electronics} A good starting point for analyzing exciton behaviour is the bandstructure of electrons and holes in CNTs, because we can express the exciton wavefunction as a linear combination of the tensor products between various electron and hole states.~\cite{many_body_quantum_Fetter_1971} If we imagine CNTs as rolled-up sheets of graphene, a natural starting point is the graphene single-electron bandstructure, which is readily computed using tight-binding methods.~\cite{graphene_tightbinding_Reich_2002,Dresselhaus_CNT_Properties} This process of rolling up the graphene sheet can be thought of as starting from some carbon atom in the lattice, moving along a vector that is a linear combination of the graphene lattice vectors, $\vec{a}_1$ and $\vec{a}_2$, to another atom, and joining these two atoms via a periodic boundary condition. The vector $n\vec{a}_1+m\vec{a}_2$ is known as the chiral vector; the ordered pair of integers $(n,m)$ is referred to as the nanotube chirality. This procedure is illustrated in Fig. \ref{Fig1}. Depending on the values of $n$ and $m$, the periodic boundary condition breaks up the graphene bandstructure into so-called cutting lines, which slice through the otherwise two-dimensional (2D) $E$--$k$ dispersion. Where these cutting lines lie (i.e., whether or not they pass through a Dirac point in the graphene bandstructure) determines whether the CNT has metallic or semiconducting properties.~\cite{Dresselhaus_CNT_Properties}

\subsection{Exciton Wavefunctions}
If we assume the electron and hole wavefunctions are known, we can write the exciton states as
\begin{equation}
    \ket{l} =  \sum_{k_c,k_v}A_l(k_c,k_v)\hat{u}^{\dagger}(k_c)\hat{v}(k_v)\ket{GS},
\end{equation}
where $l$ stands in for all exciton quantum numbers, $k_c(k_v)$ is the conduction (valence) band electron momentum, $\hat{u}^{\dagger}$ is the creation operator for an electron in the conduction band, $\hat{v}$ is the annihilation operator for an electron in the valence band, and $\ket{GS}$ is the ground state, in which there are no conduction band electrons. This expansion is referred to as the Tamm-Dancoff approximation.~\cite{many_body_quantum_Fetter_1971}

The coefficients $A_l$ are the amplitudes of the different states. The momenta $k_c$ and $k_v$ will only take the values that lie along the cutting lines.

To find the coefficients $A_l$ and the exciton energies, we solve the Bethe-Salpeter equation,
\begin{equation}
    [E_c(k_c)-E_v(k_v)]A_l(k_c,k_v) = \sum_{k_{c'},k_{v'}}\kappa(k_c,k_v;k_{c'},k_{v'})A_l(k_{c'},k_{v'}),
\end{equation}
where $E_c$ and $E_v$ are the conduction and valence band electron energies, respectively, and $\kappa$ is the interaction kernel that describes particle--particle interaction. The kernel $\kappa$ can be broken up into a direct term and an exchange term under the GW approximation.~\cite{greens_functions} The direct interaction uses the screened Coulomb potential, while the exchange interaction uses the bare Coulomb potential (for computational simplicity). The single-particle momenta, $k_c$ and $k_v$, can be used to form two two-particle momenta, $K$ and $k_r$, which are the exciton center-of-mass and relative momentum, respectively, defined as
\begin{equation}
    K = \frac{k_c-k_v}{2},\quad k_r = \frac{k_c+k_v}{2}.
\end{equation}
The center-of-mass momentum is important to consider because, owing to momentum-selection rules,~\cite{Dresselhaus_symmetry_2006_phys_rep,Dresselhaus_symmetry_2006,Dresselhaus_symmetry_1993} excitons with nonzero center-of-mass momentum cannot be optically excited. The details of this calculation can be found in Refs. [\onlinecite{Davoody_exciton_transfer,dresselhaus_CNT_exciton_TB}].

\subsection{Transfer Rates}
The process of exciton transfer between donor and acceptor tubes forms the computational backbone of the Monte Carlo simulation. These transfer processes typically have rates in the $10^{12}$--$10^{14}\, \mathrm{s}^{-1}$ range.

\begin{figure*}
\begin{center}
\includegraphics[width=\linewidth]{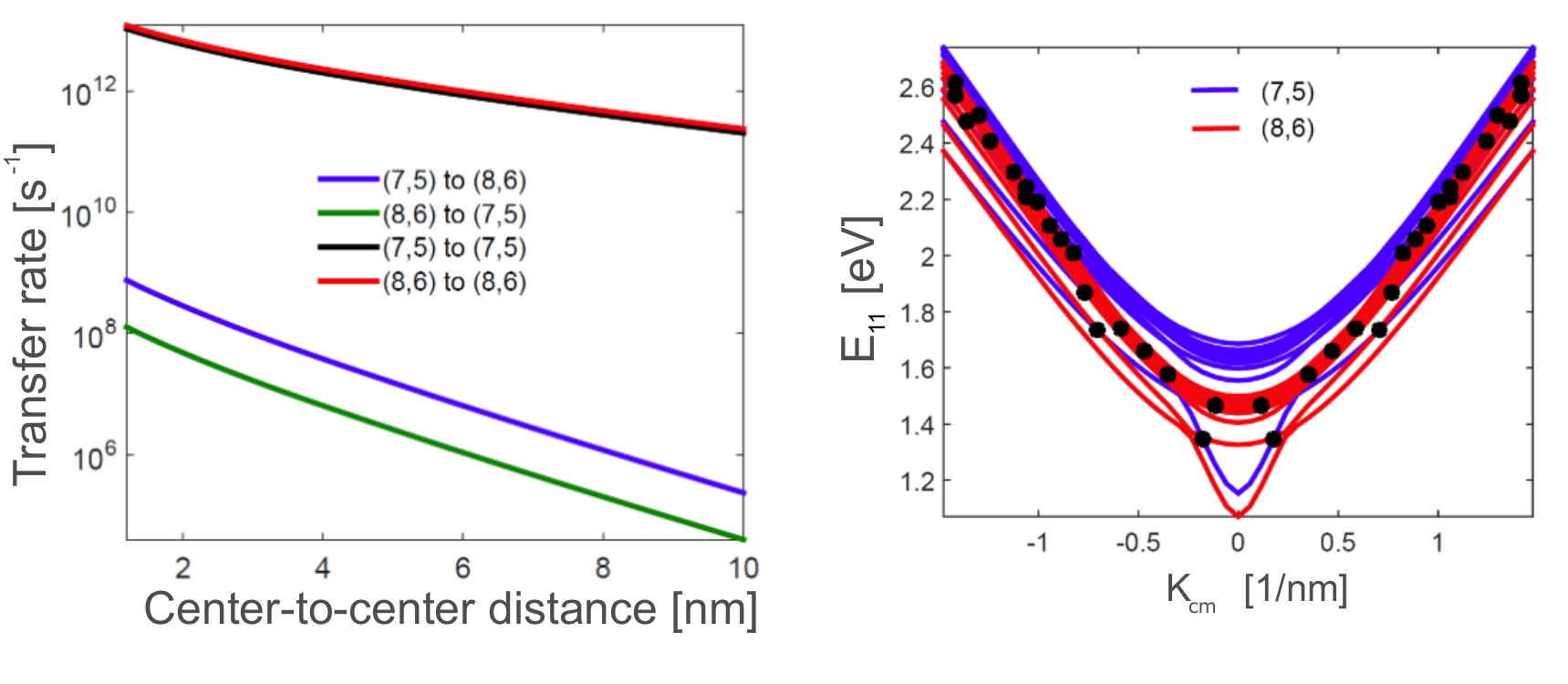}
\caption{{a) Resonant-transfer rate as a function of center-to-center distance for all possible donor/acceptor combinations of (7,5) and (8,6) CNTs. b) Bandstructure for S11 excitons in (7,5) and (8,6) CNTs. S11 refers to the exciton formed from a hole in the highest valence band and an electron in the lowest conduction band.}}\label{Fig2}
\end{center}
\end{figure*}

The transition rate for resonant exciton transfer~\cite{Jones_2019_RET_review} between state $s_1$ in tube 1 and state $s_2$ in tube 2 (these states will have the same energy) is calculated based on Fermi's golden rule as

\begin{equation}\label{eq:rate}
    \Gamma_{12} = \frac{2\pi}{\hbar}\sum_{s_1,s_2}\sum_{K_1,K_2}\frac{e^{-\beta\Omega_{s1}}}{Z}|M_d|^2\delta(\Omega_{s_1}-\Omega_{s_2}),
\end{equation}
where $Z$ is the partition function for the excitons in tube 1 and $M_{d(e)}$ is the matrix element calculated for the direct (exchange) Coulomb interaction as
\begin{equation}
    M_d = \sum_{k_{r_1}}\sum_{k_{r_2}}A^*_{s_1}(K_1,k_{r_1})A_{s_2}(K_2,k_{r_2})\bra{k_{v_2},k_{c_1}}v(|r-r'|)\ket{k_{c_2},k_{v_1}}.
\end{equation}
Here, the coefficients $A_{s_n}$ come from solving the Bethe-Salpeter equation, Eq. (2). $v(|r-r'|)$ is the screened Coulomb potential for the direct interaction,
\begin{equation}
    v(|r-r'|) = \frac{e^2}{4\pi\epsilon|r-r'|},
\end{equation}
where $\epsilon$ is the average value of the permittivity of the medium between CNTs. This value might change depending on the process used to separate or align CNTs.~ \cite{alignment_jinkins_2019,CNT_alignment_arnold,Wei_chiral_enantiomer_separation_2018,Ihara_metal_separation_2011,Sturzl_chiral_separation_2009,Yomogida_chiral_separation_2016,Graf_2016_chiral_separation}
The rates depend on the bandstructure, distance between, and relative orientation of the donor and acceptor CNTs. In Fig. \ref{Fig2}, we show an example of the bandstructure for two types of CNTs  and the resonant-transfer rates as a function of distance calculated using Fermi's golden rule for the direct interaction. In this work, we have implemented the rates based on resonance energy transfer between tubes mediated by the direct Coulomb interaction (the exchange interaction is negligible owing to tube separation) as detailed in Ref. [\onlinecite{Davoody_exciton_transfer}]. Phonon scattering is too short range to enable intertube transfer, but contributes to higher-order nonresonant processes that may involve dark excitons, as discussed, for example, in Refs. [\onlinecite{Davoody_exciton_phonon_transfer}] and [\onlinecite{Postupna_2014_JPCL}].

\section{Mesh Generation}\label{sec:mesh_generation}
The dynamics of excitons in a CNT film is important for energy-harvesting devices. Excitons are generated on one side of the CNT film and must make it to the opposite side, which is in contact with a $\mathrm{C}_{60}$ layer, in order to dissociate into a free electron and free hole capable of contributing to electrical current.~\cite{CNT_c60_interface_Dowgiallo_2016}. Excitons that dissociate before reaching the $\mathrm{C}_{60}$ layer do not contribute to the current, which is why premature dissociation needs to be minimized. Here, we simulate realistic CNT films in order to understand the limitations of exciton diffusion in CNT-based photovoltaics and improve their energy-harvesting efficiency.

\subsection{Mesh-Generation Basics}

A CNT film consists of individual CNTs, which are themselves composed of CNT segments. Film generation is accomplished at the segment level. We begin by defining a simulation domain, which in our case is a cube of dimensions 200 nm $\times$ 200 nm $\times$ 200 nm. These dimensions are chosen so that we can simulate a 5--20-nm-thick film, as reported in Refs. [\onlinecite{mehlenbacher_2016_cnt_films_spectroscopy}] and [\onlinecite{film_thickness_Shea_2013}]. We found it easiest to orient the CNTs in the $x$-$z$ plane, so cross-plane refers to movement along the $y$-direction and in-plane refers to motion in the $x$-$z$ plane. (The somewhat nonstandard choice of label --- $y$ instead of $z$ for the priority direction --- actually stems from the settings in Bullet Physics.)

With the simulation boundaries defined, we begin the process of generating CNT segments. Each CNT segment is a rigid cylinder. The length of the segment is such that the number of segments $\times$ length of segments gives the CNT length. The exact number of segments depends on the film morphology, because, for perfectly aligned CNTs, we only need a few segments. For unaligned morphologies, we used around ten segments and for the aligned morphology we used three segments for each tube. Each segment is connected to others with a cone constraint, illustrated in Fig. \ref{Fig4}(a), which allows rotation but keeps the segments connected. In the extreme case of a single segment per CNT, there would be no variation in the orientation of the CNT along its length, i.e., the tube would be a single rigid cylinder not capable of bending, and would require only a few numbers (coordinates and orientation) to represent. In the other extreme of a continuum of segments, there could be an arbitrarily large number of bends in the CNT, but an arbitrarily large set of numbers would be required to represent each point on the CNT (many coordinates and orientations). We found ten segments per CNT to be a reasonable compromise between realistic mesh morphology and computational restrictions. The CNT length can be fixed or chosen from a distribution. Section \ref{sec:mesh_options} describes the selection of CNT length in more detail. The segment diameter is determined from CNT chirality.

Each CNT segment is given a position (the $(x,y,z)$ coordinates of its center) and an orientation. For generation purposes, each segment within a CNT has the same orientation but collisions with other CNTs later on will add some randomness. The orientation is defined by picking an angle $\theta$ between $0$ and $\pi$, and defining an orientation vector, $\hat{v} = \cos{\theta}\hat{x}+\sin{\theta}\hat{z}$, that points along the axis of the tube segment. The angle $\theta$ can be randomly selected from a uniform distribution to create a film consisting of randomly oriented CNTs, or can be given a specific value to create films of aligned CNTs. Segments can rotate relative to one another but must obey the cone constraint. For example, the second tube segment cannot rotate in such a way that it overlaps with the first, occupying the same volume. Similarly, the second tube segment cannot be arbitrarily far away; it must remain attached at one endpoint to the first segment. This process of segment generation continues until the CNT has reached a predetermined final length.

Once a group of CNTs is complete, they are released and allowed to fall under the influence of gravity to the bottom of the container. The Bullet Physics software simulates collisions, so that the CNTs bounce off each other elastically. The dynamics of CNTs as they fall and collide is governed by Newton's laws of motion. Eventually, there will be hundreds of CNTs on the bottom of the rectangular container, and the computational burden of simulating the dynamics of every single CNT that has already fallen will become too great and also unnecessary. To circumvent this, we ``freeze'' CNTs in place after enough have been generated, so that at most 70 are still moving at a time. This number can be changed to suit slower/faster machines than we had access to. Also, it should be noted that the number of CNTs active should be larger than the number of CNTs added together as discussed in Sec. \ref{sec:mesh_options}, because otherwise some of the CNTs added to the simulation will freeze before they are allowed to fall and settle. These frozen CNTs no longer participate in the collision dynamics of the film, but will still be part of the exciton Monte Carlo simulation. An example of the CNT aggregate generated in Bullet Physics can be seen in Fig. \ref{Fig4}(c).

\begin{figure}
\begin{center}
\includegraphics[width=\linewidth]{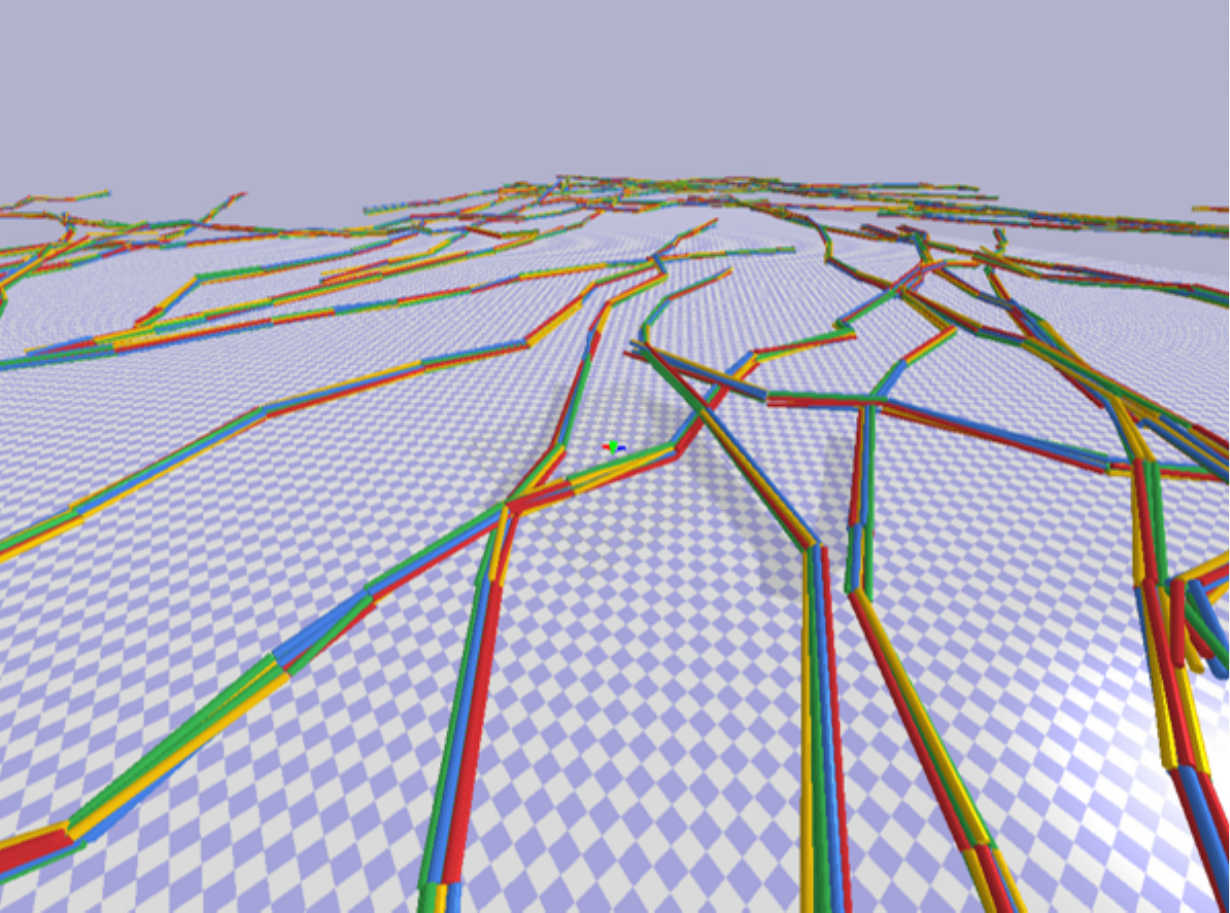}
\caption{Representative image of a CNT mesh during the generation process. In this image, the bundling constraint has been applied to create hexagonally bundled CNT fibers.}\label{Fig3}
\end{center}
\end{figure}

\begin{figure}
\begin{center}
\includegraphics[width=\linewidth]{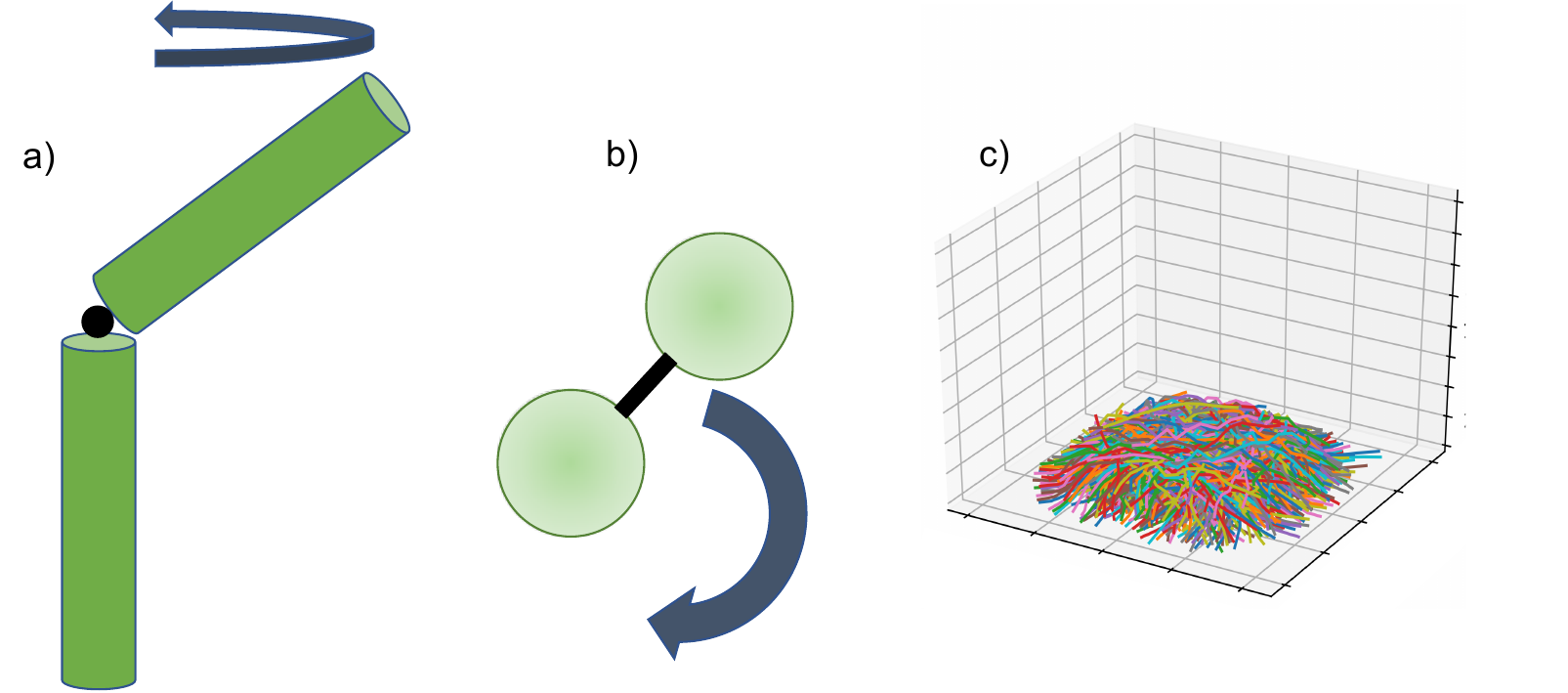}
\caption{(a) Cone constraint between different segments within the same CNT. (b) Bundling constraint between CNTs within the same hexagonal bundle. (c) Example of a film generated using Bullet Physics.}\label{Fig4}
\end{center}
\end{figure}

\subsection{Mesh-Generation Options}\label{sec:mesh_options}

Some parameters related to mesh generation were briefly mentioned in the previous section. Here we discuss all options in more detail and give examples of situations where different choices would make sense given the physical problem.

\begin{enumerate}
    \item Visualisation: A graphical representation of the CNTs can be displayed during the mesh generation. This carries a significant computational burden, but is useful for debugging.

    \item Simulation domain size: The square box representing the simulation domain can be made any desired size. We chose 200 nm because, with a film thickness of 10 nm, typical in experiment,  ~\cite{film_thickness_Shea_2013} this allows enough space for excitons to diffuse a significant distance before hitting the boundary.

    \item Drop height: The height from which CNTs are dropped. In order to prevent CNTs falling straight through the simulation boundary, this height should be greater than $\frac{1}{2}g(\Delta t)^2$, where $g$ is gravitational acceleration and $\Delta t$ is the time step. We used 20 nm.

    \item Intertube spacing: Additional spacing between CNTs, beyond the center-to-center distance stemming from finite CNT diameter.

    \item CNT length: Length of an individual CNT. This parameter can be provided as a range, with each CNT having a length selected from a distribution, such as the lognormal distribution reported in Ref. [\onlinecite{CNT_length_Streit_2012}]. Default is uniform distribution. For the data in this paper, we set all lengths to 100 nm because nothing in our simulations was sensitive to individual CNT length.

    \item CNT section length: Length of a single section within a CNT. This parameter can be provided as a range in the form $[L_{min},L_{max}]$. We used a range of 7--12 nm and added segments to get as close to 100-nm length as possible without going over.

    \item Number of tubes added together: Number of CNTs fully generated before the group is dropped and allowed to settle. This number should be less than the next option, Number of active tubes, to avoid CNTs freezing before they are dropped.

    \item Number of active tubes: Maximum number of CNTs that can remain unfrozen in the simulation. We use 70, but this number depends greatly on the specifications of the particular machine(s) being used for the mesh generation.
        \setcounter{saveenum}{\value{enumi}}
\end{enumerate}

Beyond these options, there are two other interesting capabilities that require somewhat more involved changes to the code, but are useful for specific physical situations. These are CNT alignment and bundling.

\begin{enumerate}
 \setcounter{enumi}{\value{saveenum}}
\item Instead of choosing random orientations for CNT segments, each tube can be oriented in a desired direction, such as along the $x$ or $z$ axes. This type of alignment~\cite{alignment_jinkins_2019,cantoro_alignment_2006,CNT_alignment_arnold} is preferred for certain applications,~\cite{Jinkins_2019} such as in thin-film transistors.~\cite{Vaillancourt_2008_CNT_transistor,Xiao_2003_transistor} The alignment is accomplished by simply defining the angle $\theta$ discussed above to be a fixed value, rather than selecting it randomly for each CNT segment.

\item Bundling involves adding another constraint to the segments, joining the segments of several CNTs together in such a way that they are allowed to rotate around one central CNT, but are kept at a fixed distance from it. This can be done with seven CNT segments, one central and six surrounding, to mimic the hexagonally packed morphologies described in Ref. [\onlinecite{hexagonal_CNT_bundles_Grechko_2014}].
\end{enumerate}

\subsection{Mesh Interpolation}
\begin{figure}
\begin{center}
\includegraphics[width=\linewidth]{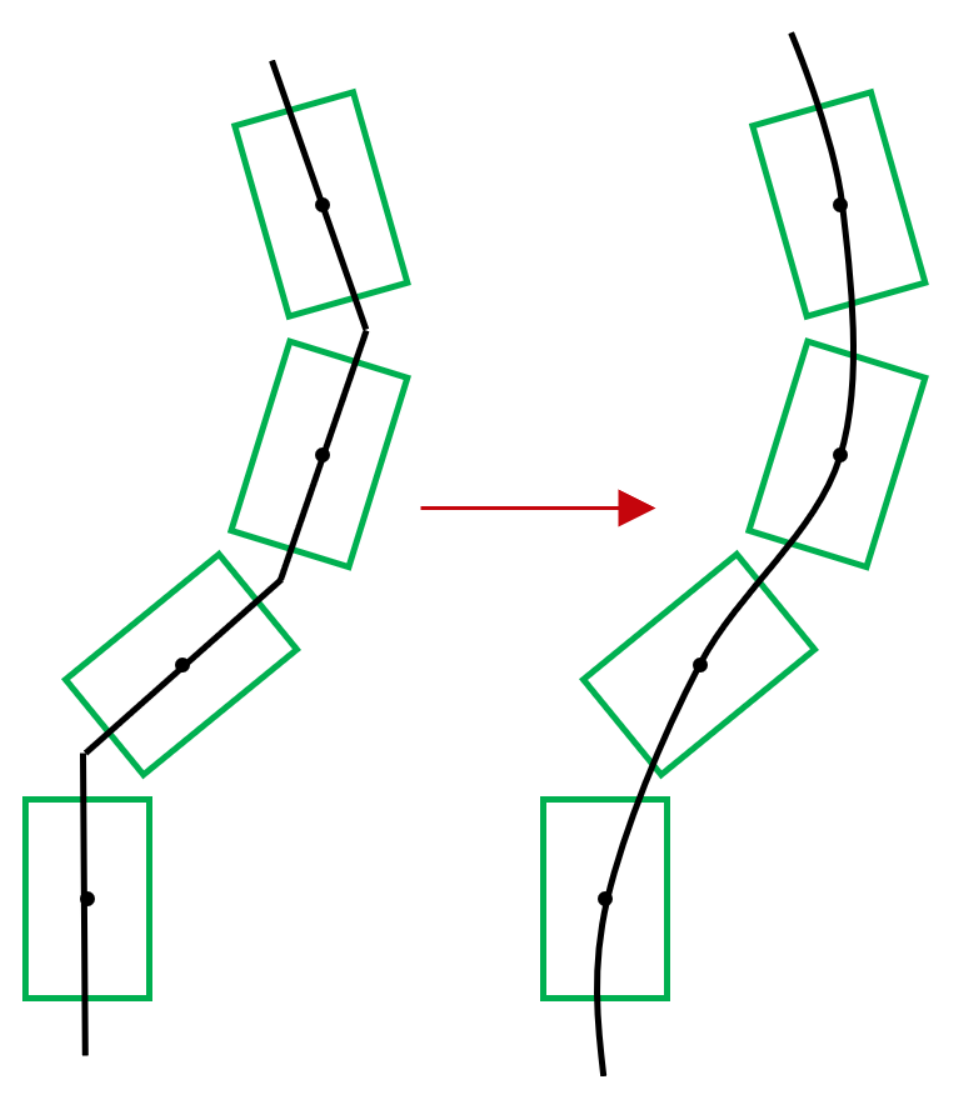}
\caption{When the mesh is generated using Bullet Physics, only the center coordinate of each segment is saved. We use B-spline interpolation through Python's Scipy module to increase the density of points along each CNT.}\label{Fig5}
\end{center}
\end{figure}

The process described in the previous two sections results in the following: the $(x,y,z)$ coordinates of the center of each CNT segment; two angles describing each CNT segment's orientation relative to the axes of the global Cartesian coordinate system; the length of each CNT segment; the diameter of each CNT segment; and the chirality of each CNT segment. We would like a higher density of points per unit length along each CNT to more accurately model exciton motion, but making the segments smaller drastically increases the computation time required for mesh generation. For reference, mesh generation took about twenty-four hours on one of our machines, which has the following specifications: Intel Core i7-8700 CPU at 3.20 GHz, 3192 MHz, 6 cores, 12 logical processors, and 32 GB of memory. Instead, we take our mesh generated using Bullet, and input it into a separate Python script. In this script, we figure out which mesh points belong to each CNT. For all points within a single CNT, we interpolate using a third-order polynomial to increase the density of points along the CNT length to 100. This process (Fig. \ref{Fig5}) is repeated for all CNTs in the film, and we produce an output file that contains all the same data that we mentioned above, only with a finer mesh density along each CNT. The interpolation is accomplished using the interpolate function from Python's popular Scipy module,~\cite{Scipy} with adjustable parameters controlling the polynomial order and smoothness used for interpolation. For all data presented in this paper, we used a polynomial order of three and smoothness of 500.

\subsection{Intertube Spacing}
The center-to-center distance between two CNTs is determined by the CNT diameter, which, in turn, depends on CNT chirality. However, several of the chemical processes used to select certain chiralities and align CNTs result in the deposition of a polymer coating on the CNTs, ~\cite{Yomogida_chiral_separation_2016,Wei_chiral_enantiomer_separation_2018,Sturzl_chiral_separation_2009,Ihara_metal_separation_2011,CNT_alignment_arnold,alignment_jinkins_2019,cantoro_alignment_2006,Graf_2016_chiral_separation} which increases the separation between CNTs in a film. We can reproduce this effect by adding a fixed value to the diameter of each CNT. We expect a significant effect from even small additional intertube spacing, because the transfer rates are second order in the matrix element that couples initial and final states.~\cite{Davoody_exciton_transfer}

\subsection{Quenching Sites}
The processes used to align CNTs and select certain chiralities \cite{Yomogida_chiral_separation_2016,Wei_chiral_enantiomer_separation_2018,Sturzl_chiral_separation_2009,Ihara_metal_separation_2011,CNT_alignment_arnold,alignment_jinkins_2019,cantoro_alignment_2006} may introduce defects that act as trapping sites for excitons. If the exciton encounters one of these sites, it may dissociate or become trapped, unable to traverse the film.

These quenching sites introduce trap states that have energies below the band minimum, and are narrowly localized in real space (delocalized in the reciprocal $k$-space). Any exciton whose wavefunction overlaps sufficiently with a quenching site will jump into the trap state. The likelihood of jumping back out is very small, so we can say that the exciton remains trapped for the duration of the simulation. Since we already must choose a scattering site for each exciton in order to employ our transfer rates, we can monitor how close each exciton is from these quenching sites, and simulate the trapping of an exciton by setting all transfer rates to zero once it gets too close to a quenching site, which will prevent the exciton from leaving. We have two tuning parameters for the quenching sites. The first one is the size, which is the radius around the quenching site within which an exciton becomes trapped (i.e., has all scattering rates set to zero and stops moving). The second parameter is the percentage of scattering sites that will act as quenching sites. For now, we use 2 nm as the quenching-site radius and vary the number of scattering sites, as seen in Sec. \ref{sec:Results}.

\section{Ensemble Monte Carlo}\label{sec:EMC}

The EMC method is a stochastic technique used for solving differential and integro-differential equations.~\cite{Jacoboni_1983_EMC_rev} Here, we use it to solve the exciton equation of motion. Excitons ``hop'' from site to site, rather than move continuously (optically generated excitons have zero center-of-mass momentum and move only when they scatter). We can write the rate equation for the population of each site as
\begin{equation}\label{eq:rate_eqn}
    \frac{dN_i}{dt} = -\sum_{j}N_i\Gamma_{ij}+\sum_{j}N_j\Gamma_{ji},
\end{equation}
where $N_i$ is the population at site $i$. $\Gamma_{ij}$ is the transfer rate from initial site $i$ to final site $j$, calculated based on Eq. (\ref{eq:rate}). More details on the calculation of the scattering rates and their dependence on tube chirality, intertube distance, and orientation can be found in Refs. [\onlinecite{Davoody_exciton_transfer,Davoody_exciton_phonon_transfer}]. The dominant intertube hopping mechanism is resonance energy transfer, where the initial and final states have the same energy, and is mediated by the direct Coulomb interaction (exchange interaction is negligible owing to tube separation).~\cite{Davoody_exciton_transfer}

\begin{figure}
\begin{center}
\includegraphics[width=\linewidth]{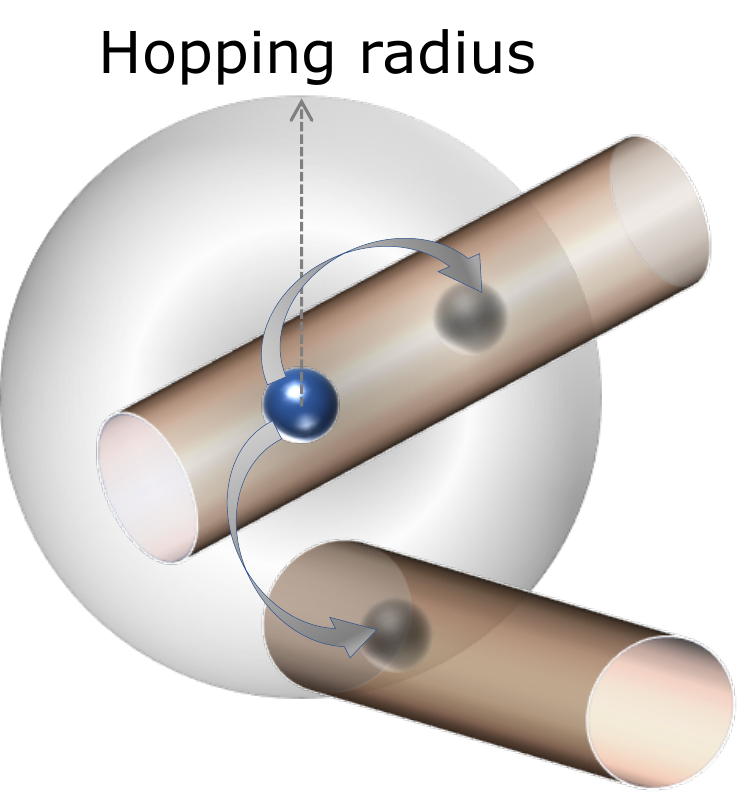}
\caption{Illustration of the scattering process and how the initial site's hopping radius is used to determine the final site after scattering. Each tube segment has a position and orientation that is stored. The segments are broken up into multiple scattering sites. Excitons can scatter to a site within the hopping radius of the initial site, with transfer rates determined by distance, relative orientation, and chirality of the donor/acceptor CNT. The image illustrates intertube and intratube possibilities for exciton hopping.}\label{Fig6}
\end{center}
\end{figure}

Each exciton is injected in the center of the mesh. Because the transfer rates depend on a screened Coulomb potential term, they fall off rapidly with distance.~\cite{Davoody_exciton_transfer} We therefore set a maximum hopping radius for excitons at each scattering site. The exciton experiences a series of scattering events during the simulation. The time until the next scattering event is chosen based on the scattering rates for each of the scattering sites within the current site's hopping radius. A schematic of the hopping process is shown in Fig. \ref{Fig6}.

At the simulation-domain edge, we can impose different boundary conditions, such as absorbing the exciton by removing it from the list of particles or reinjecting it into the mesh. A periodic boundary condition could be implemented by reinjecting excitons into corresponding sites on the opposite face of the simulation domain. Owing to the randomness of CNT orientations, a truly periodic boundary condition does not make much sense because the randomly oriented CNT film would just be copied in adjacent unit cells, likely not matching up at the boundary. Instead, when an exciton reaches the simulation boundary, we simply reinject it into a random CNT of the same chirality. (As discussed later in this section, since the diffusion tensor is calculated using displacements. we make sure to remove the displacement added due to the reinjection from the calculation. This way, excitons can achieve displacements greater than the simulation--domain dimensions.)

We track each exciton's positions in the mesh over time in order to calculate the diffusion tensor describing motion through the film. The diffusion tensor is calculated as \cite{diffusion_tensor_1979,diffusion_tensor_Rengel_2013}
\begin{equation}\label{eq:diff_coeff}
    D_{ij} = \lim_{t\to\infty} \frac{<x_ix_j>-<x_i><x_j>}{2t},
\end{equation}
where $t$ denotes time, $<...>$ indicates an average over the particle ensemble, and the subscripts $i$ and $j$ take vales 1,2, and 3 that correspond to the $x$, $y$, and $z$ displacements from the initial position, respectively.

Below are the options available when running the Monte Carlo simulation.

\begin{enumerate}

    \item Axis shift: Amount of space in each dimension to discretize for the purpose of calculating scattering rates. Entered in the form $[x_{i,min},x_{i,max},n]$ where $n$ is the number of points and $x_{1,2,3}$ is $x$, $y$, or $z$. The difference between $x_{min}$ and $x_{max}$ should be equal to the hopping radius.

    \item Number of particles in the simulation: In this simulation, excitons are non-interacting, but simulating a large number reduces statistical noise and opens up parallelization opportunities.

    \item CNT chirality: This parameter is used to look up the correct scattering table for excitons in the simulation. It should match up with the chirality used in mesh generation, but technically does not have to.


    \item Density of quenching sites: Density of quenching sites as a percentage of all scattering sites.

    \item Exciton velocity: Related to exciton center-of-mass momentum. It will be zero for optically excited excitons due to momentum-selection rules, but could in principle be nonzero.

    \item Keep old results: Boolean parameter. If true, the Monte Carlo code will create a new output folder. If false, and a folder with the same name as the folder the Monte Carlo code is trying to save output to exists, previous results in that folder will be overwritten.

    \item Max dissolving radius: Radius of quenching site. Excitons within this range will have all of their transfer rates set to zero.

    \item Max hopping radius: Maximum distance an exciton could move in one scattering event. This parameter is needed to discretize scattering rates. It should be large to avoid missing possible scattering events. All data presented in this paper uses 20 nm for this parameter. As shown in Fig. \ref{Fig2}(a), the transfer rate drops by two to three orders of magnitude when CNT separation increases by 10 nm, so the probability of jumps longer than 20 nm is exceedingly small. The diminishing benefits of larger hopping radii  when must be weighed against the detriment of increasing memory cost.

    \item Maximum simulation time: Simulation runtime. We found 1 ps to be long enough for the diffusion-tensor elements [Eq. (\ref{eq:diff_coeff})] to reach their steady-state values. This time is shorter than the exciton lifetimes reported by [\onlinecite{flach_2020_improved_exciton_liftime}], so we can reasonably assume that excitons do not spontaneously dissociate during our simulation.

    \item Monte Carlo time step: Time step at which simulation saves position/displacement of all excitons. We used 1 fs. This parameter in general depends on typical scattering rates.

    \item Number of sections for injection region: Number of sections in each dimension for the simulation domain. This number is used to determine the size of the injection region. In the simulation, excitons are initially created in the injection region, which is always the central section. All data presented in this paper uses 3 for this parameter, which results in an injection region that has 1/27 of the total simulation-domain volume.

    \item Rate type: Determines how transfer rates are calculated. ``Davoody'' refers to the calculations based on our previous work.~\cite{Davoody_exciton_phonon_transfer,Davoody_exciton_transfer}

    \item Temperature: All data presented in this paper is at 300 K.

    \item Theta: Number of discrete angles to resolve, $n$, entered as an array, $[\theta_{min},\theta_{max},n]$.

    \item Trim limits: Array of limits to trim domain if necessary, entered as $[x_{min},x_{max}]$
\end{enumerate}

These input options allow for fine control of the simulation in all aspects.

\section{Results}\label{sec:Results}
\begin{figure*}
\begin{center}
\includegraphics[width=\linewidth]{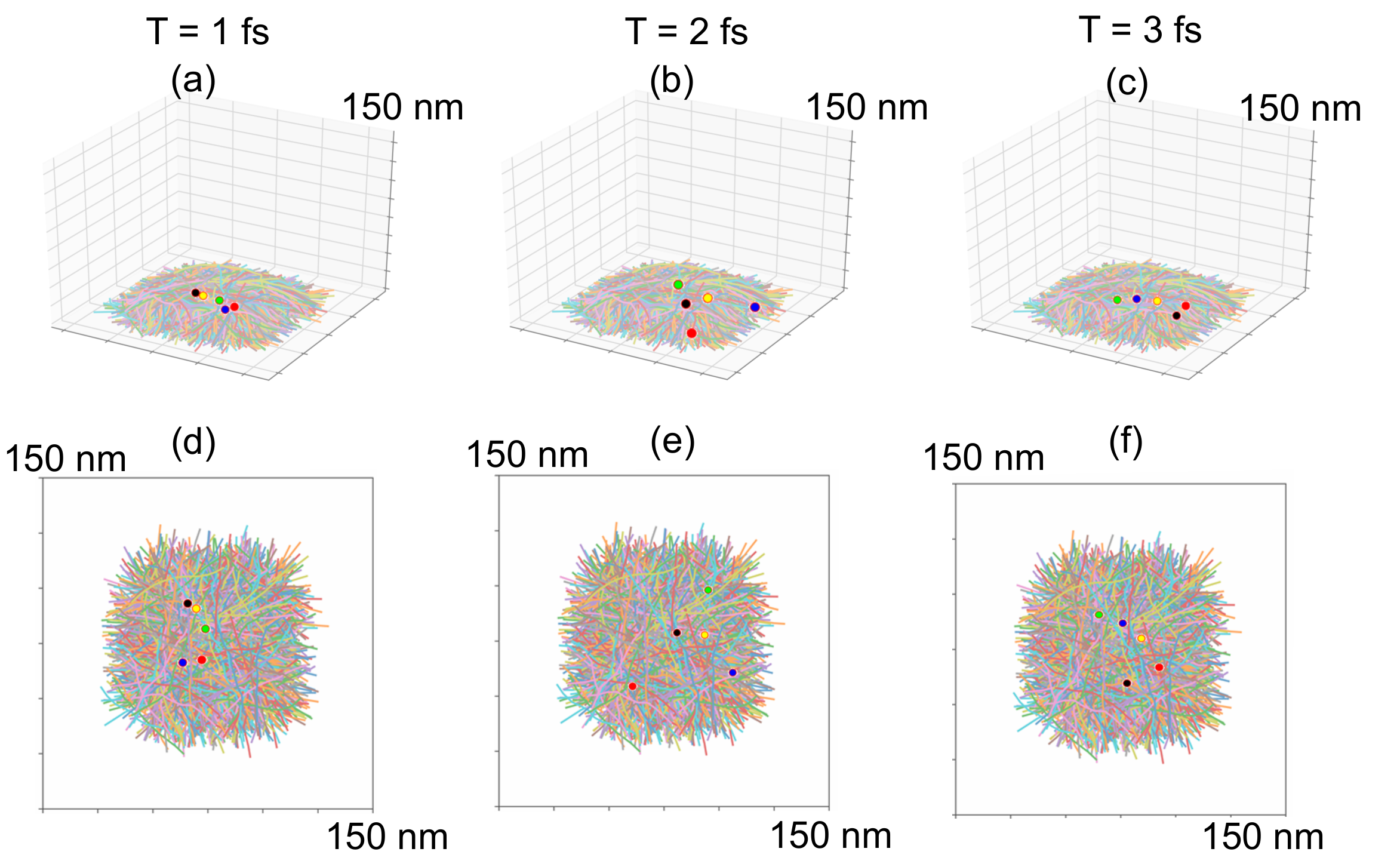}
\caption{Isometric view of position of several excitons (colored dots) within the film at three simulation time steps: (a) 1 fs, (b) 2 fs, and (c) 3 fs. (d)-(f) The same, but from top-down perspective. Excitons that leave the simulation boundary (red dot between 2 fs and 3 fs) are reinjected into the film.}\label{Fig7}
\end{center}
\end{figure*}

We run the simulation described above and tweak the film morphology, intertube spacing, and quenching-site-density to study the effects on the diffusion tensor. Each simulation domain is roughly 200 nm $\times$ 200 nm $\times$ 20 nm, where the smallest dimension is the film thickness. In Fig. \ref{Fig7}, we show the time evolution of exciton position within the film. Colored dots represent excitons, which start in the center of the simulation domain (injection region) and diffuse through the film.

We study two different chiralities, (4,2) and (6,1). They have similar diameters, which allows us to create films with similar numbers of tubes for the same simulation-domain dimensions, but their diffusion-tensor elements differ by about an order of magnitude.

\subsection{Full Diffusion Tensor}
\begin{figure*}
\begin{center}
\includegraphics[width=\linewidth]{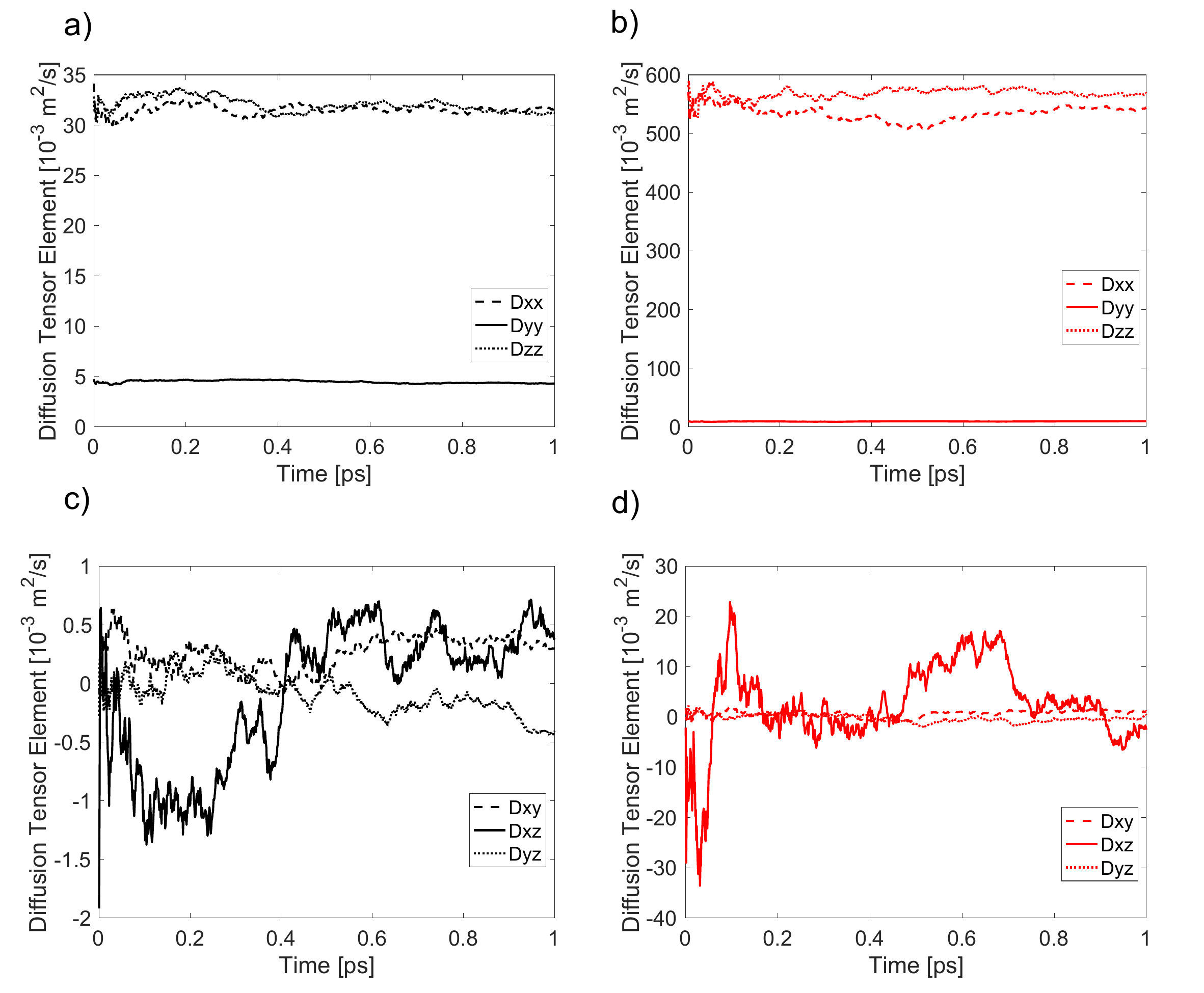}
\caption{Full two-position correlation functions versus simulation time for unbundled randomly oriented CNTs. The diffusion-tensor components are the long-time limits of these quantities. The tensor is symmetric so only the six independent components (three diagonal and three independent off-diagonal ones) are shown. Off-diagonal elements are about an order of magnitude smaller than diagonal elements, and the cross-plane element $D_{yy}$ differs from the in-plane elements $D_{xx},D_{zz}$ because, in general, transport within a tube occurs at a higher rate than transport between two tubes owing to orientation dependence. Panels (a) and (b) correspond to the diagonal terms for (4,2) and (6,1) chirality CNT films, respectively, while (c) and (d) correspond to the same, but for the off-diagonal terms.}\label{Fig8}
\end{center}
\end{figure*}

First, we calculate the two-position correlation functions as a function of time for (4,2) and (6,1) CNTs for a single morphology consisting of single CNTs randomly oriented in Fig. \ref{Fig8}. Henceforth, we will refer to this morphology as ``single random.'' The two other morphologies are hexagonally bundled randomly oriented CNTs, ``bundled random'', and parallel CNTs, ``parallel''. The diffusion tensor is calculated by taking the long-time limit of the two-position correlation functions, as per Eq. (\ref{eq:diff_coeff}). The tensor is symmetric, so we show the three diagonal and three independent off-diagonal elements. By the time the simulation has finished, the tensor elements have reached their steady-state values.

We note two important features in Fig. \ref{Fig8}. First, all off-diagonal elements are about one order of magnitude smaller than the smallest (i.e., cross-plane) diagonal element, meaning there is little correlation between movement in different directions. Second, the cross-plane diffusion tensor element ($D_{yy}$) is considerably smaller than the in-plane diffusion tensor elements ($D_{xx},D_{zz}$), because the transfer rate for intratube movement is generally higher than that for intertube movement owing to orientation dependence.~\cite{Davoody_exciton_transfer} As we will see below, this difference is greatly reduced or vanishes for the parallel morphology, where orientation is removed as a degree of freedom.

\subsection{Morphology}
\begin{figure*}
\begin{center}
\includegraphics[width=\linewidth]{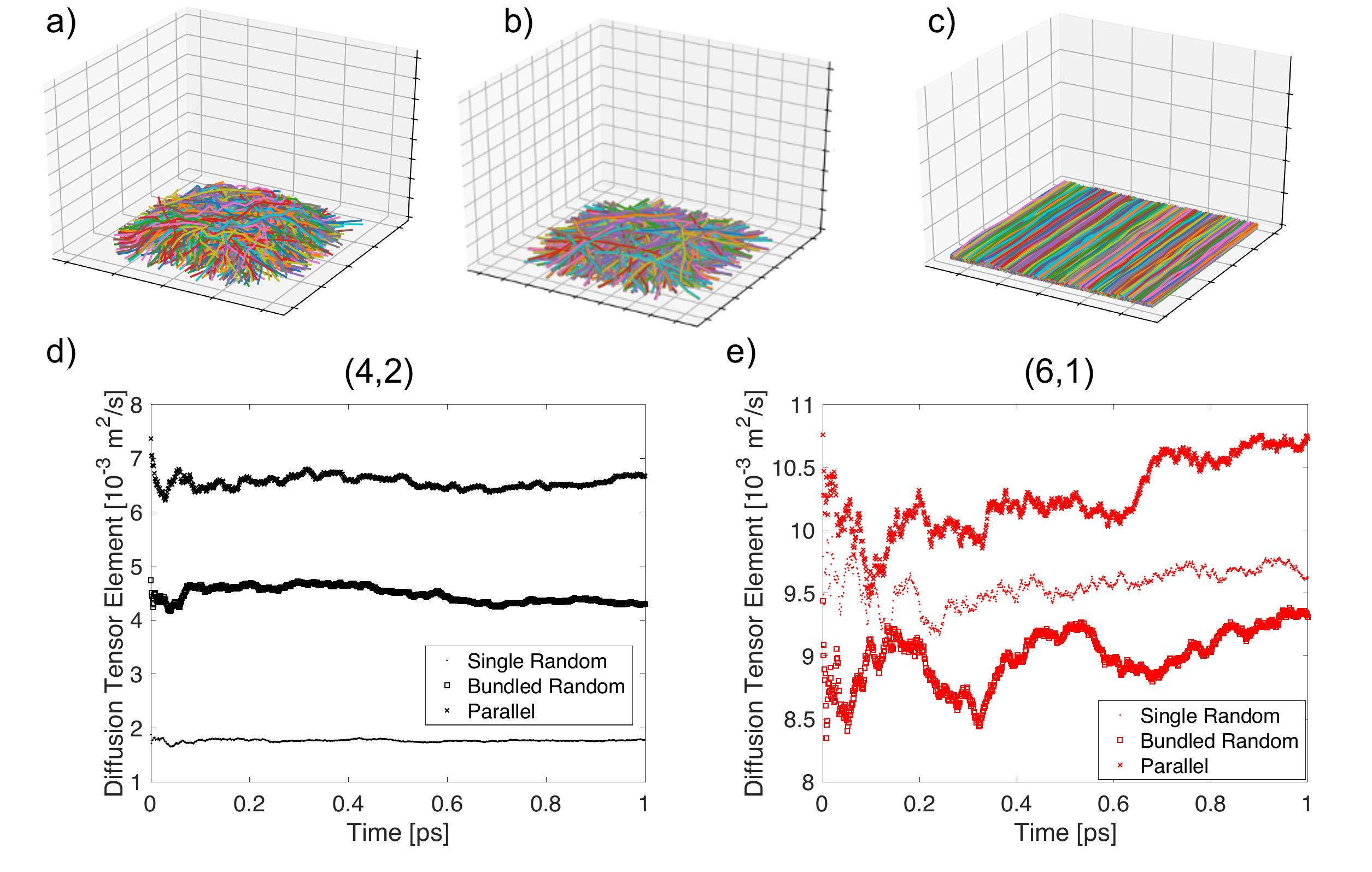}
\caption{(a)--(c) Three film morphologies: (a) single random, (b) bundled random, and (c) parallel tubes. (d) and (e) Position--position correlation function versus time for two chiralities, (4,2) in (d) and (6,1) in (e), and each of the three morphologies depicted in the top row. For both chiralities, the parallel morphology increases cross-plane diffusion by eliminating tube orientation as a degree of freedom in the transfer-rate calculation.}\label{Fig9}
\end{center}
\end{figure*}

Next we study the impact that film morphology has on the cross-plane diffusion tensor element. We focus on this element because it is the most interesting one for devices utilizing CNT films, where the excitons are excited on one side and must travel through the film thickness to a harvesting layer. Our three morphologies are single random, bundled random, and parallel, and we again study films consisting of (4,2) and (6,1) CNTs. The results can be seen in Fig. \ref{Fig9}.

There is an obvious increase in cross-plane diffusion for the parallel morphology because the CNTs involved in any transfer interaction are guaranteed to be parallel, removing orientation as a degree of freedom. A similar, but smaller effect is seen in the bundled random morphology for (4,2) CNTs, where CNTs within a bundle will always have parallel partners to transfer to, but interbundle transfer once again introduces orientation to the rate calculations. The magnitude of both of these effects depends on chirality, as demonstrated by the drastic difference in impact that morphology has on (6,1) and (4,2) films. Such a large effect indicates that experimental efforts to align tubes are worthwhile for films consisting of certain chiralities.

\begin{figure}
\begin{center}
\includegraphics[width=\linewidth]{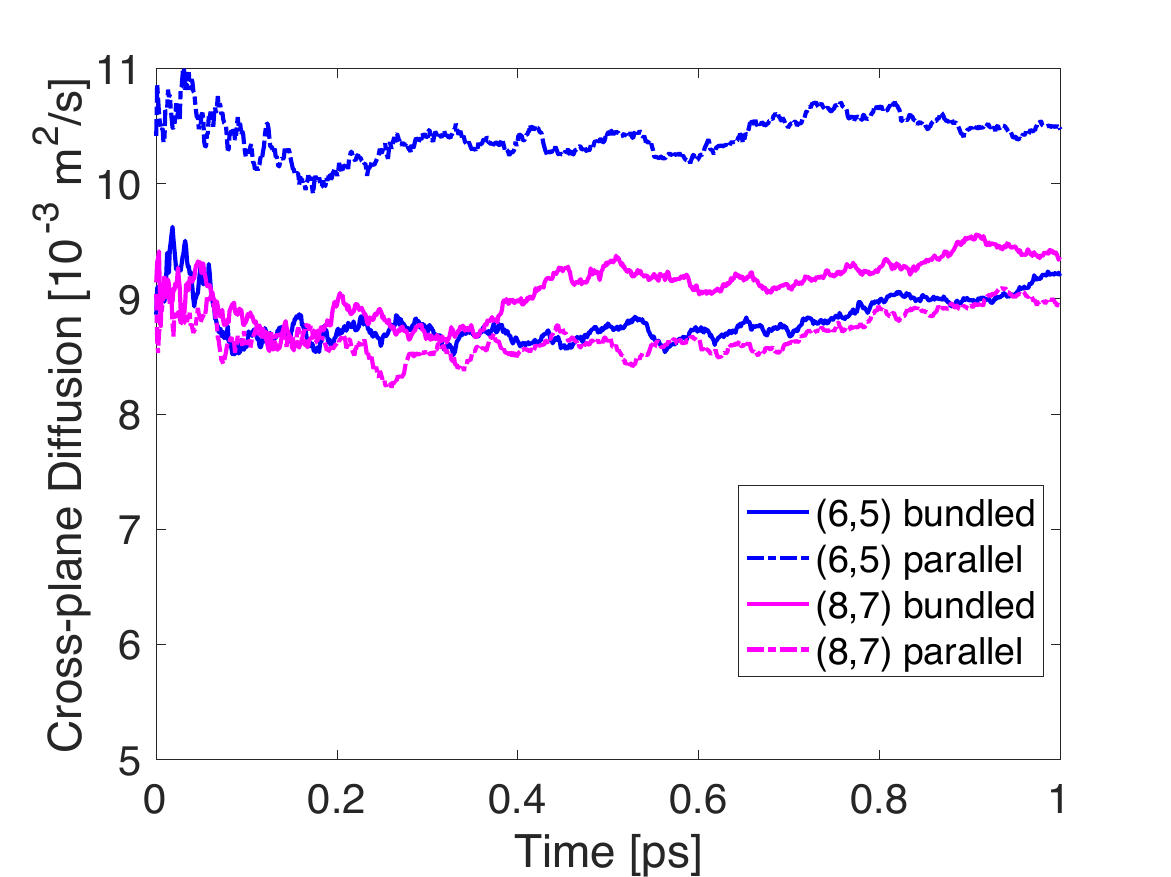}
\caption{Cross-plane diffusion tensor element as a function of time for films of (6,5) and (8,7) CNTs. We see only a small difference between the parallel and bundled morphologies, indicating that CNT size dominates over orientation as a degree of freedom in the transfer rates.}\label{Fig10}
\end{center}
\end{figure}

CNT size plays a significant role in the impact of morphology because transfer rates depend so strongly on distance. We found that, in all the chiralities we studied, (4,2) CNTs have diffusion properties that depend the most on morphology. For all other chiralities in this study [(6,1),(6,5), and (8,7)], we see only a small difference in cross-plane diffusion with different film morphologies, because the tubes are large enough that the center-to-center distance between two scattering sites dominates over other parameters, such as orientation. In Fig. \ref{Fig10}, we show a comparison between the bundled and parallel CNT morphologies for films of (6,5) and (8,7) CNTs, both of which are larger than (4,2) and (6,1) CNTs to further illustrate this dominance of size in transport properties.

\subsection{Intertube Spacing}
\begin{figure*}
\begin{center}
\includegraphics[width=\linewidth]{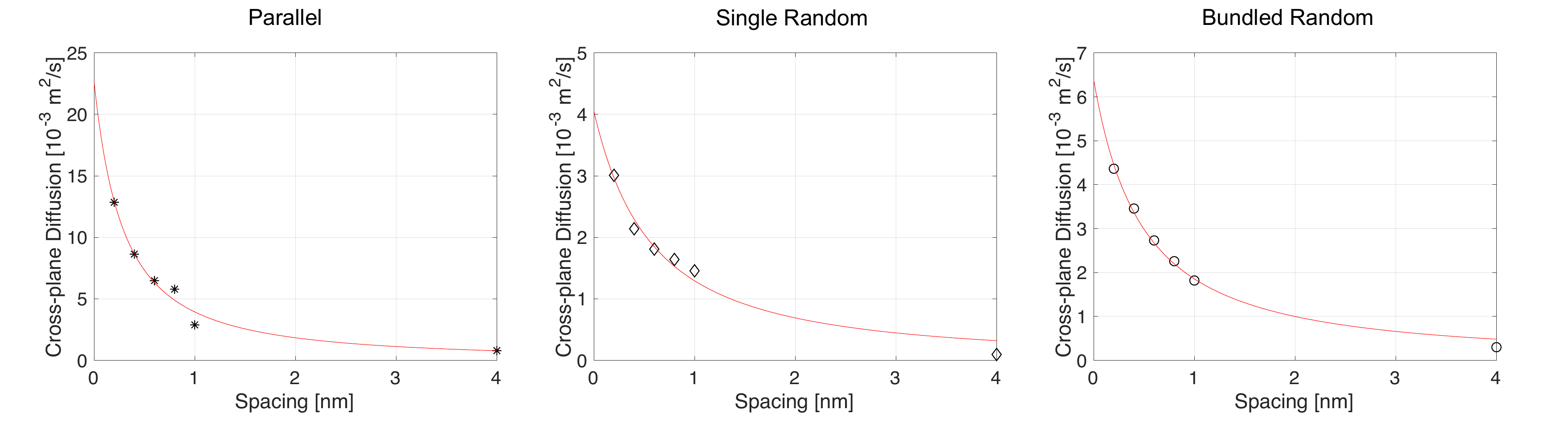}
\caption{Cross-plane diffusion-tensor element as a function of intertube spacing. Diffusion decreases with increased spacing for all morphologies, but the rate of this decrease depends on morphology. The parallel morphology exhibits the quickest decrease, while the bundled random film exhibits the slowest. This is because CNTs with random orientation could be partially oriented in the cross-plane direction and CNTs within a bundle don't have added spacing, allowing for motion in the cross-plane direction within a given bundle. The red lines are numerical fits according to $D(0)/(\underline{d}+d_0)^n$, where $\underline{d}$ is the value (in nm) of the wall-to-wall spacing (horizontal axis in all graphs). $d_0$ is no less than the tube diameter (in nanometers) and $n<2$, both dependent on film morphology and average segment orientation. Fit parameters: (parallel) $D(0)= 6.4\times 10^{-3}\,\mathrm{\mathrm{m^2/s}}$; $d_0= 0.41$ (exact tube diameter in nm), $n = 1.4$; (single random) $D(0) = 2.3\times 10^{-3}\, \mathrm{m^2/s}$; $d_0 = 0.62$; $n = 1.2$; (bundled) $D(0) = 3.5\times 10^{-3}\, \mathrm{m^2/s}$; $d_0 = 0.62$; $n = 1.3$.}\label{Fig11}
\end{center}
\end{figure*}

We now consider a single chirality, (4,2), and examine the effect of added intertube spacing on the cross-plane diffusion for our three morphologies. Processes used to select semiconducting CNTs from a solution often use a polymer~\cite{Arnold_Films_Harvesting_2011,polymer_wrapping_Nish_2007} that can be difficult to remove afterward. This polymer layer will increase the distance between CNTs in a film, which has a direct, and large, effect on the transfer rate. The results are shown in Fig. \ref{Fig11}. The red curves in Fig. \ref{Fig11} are least-squares fits of the form $D(0)/(\underline{d}+d_0)^n$, where $\underline{d}$ is value (in nm) of the wall-to-wall spacing (horizontal axis in all graphs). $d_0$ is no less than the tube diameter (in nanometers), with the exact value dependent on film morphology and average segment orientation. $n<2$ is the power characterizing the decay with increasing tube separation. (The scattering potential is Coulomb in origin and the rates (and diffusion constant) are proportional to the square of the potential matrix elements. Between pointlike objects, the Coulomb potential decays as the inverse of distance, but for an infinite charged cylinder the potential would decay more slowly (logarithmically). This is why, for the mixture of tube segments of different lengths and orientations, a decay of the potential is likely somewhat slower than the inverse of distance, and consequently the rates (and the diffusion constant) would decay more slowly than the inverse of distance squared. Hence the expectation $n<2$, as shown in Fig. \ref{Fig11}, with the exact power being dependent on the film morphology.)

The cross-plane diffusion coefficient decreases with increasing intertube spacing for all morphologies, and does so fastest in the parallel morphology. The single random morphology has a slower decay curve because it is possible for individual CNTs to be partially oriented with the cross-plane direction. Therefore, even when intertube transfer is impossible, there is some movement in the cross-plane direction. In the bundled random morphology, the orientation effect is coupled with the lack of spacing between CNTs in the same bundle, which still allows intrabundle transfer. This points to an interplay between the polymer wrapping and alignment: While alignment is good for cross-plane diffusion, it also causes the greatest sensitivity to intertube spacing. There will be a point at which there is too much intertube spacing for alignment to be worthwhile.

\begin{figure}
\begin{center}
\includegraphics[width=\linewidth]{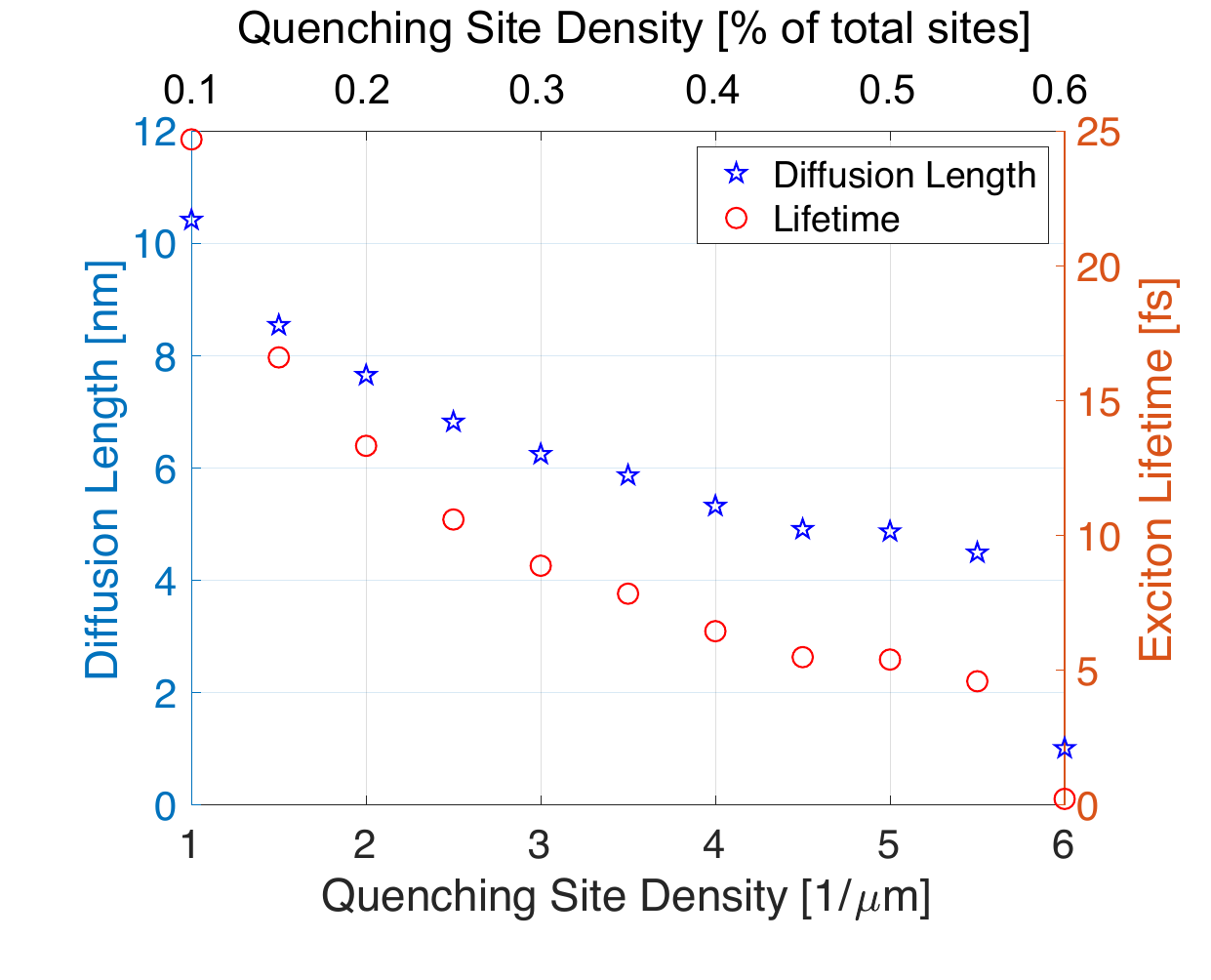}
\caption{Cross-plane diffusion length (left axis) and exciton lifetime (right axis) vs. quenching-site density in absolute units (bottom axis) and as a percentage (top axis). We see a sharp drop in both quantities beyond quenching-site density of $0.55\%$ because the exciton lifetime drops below one simulation time step.}\label{Fig12}
\includegraphics[width=\linewidth]{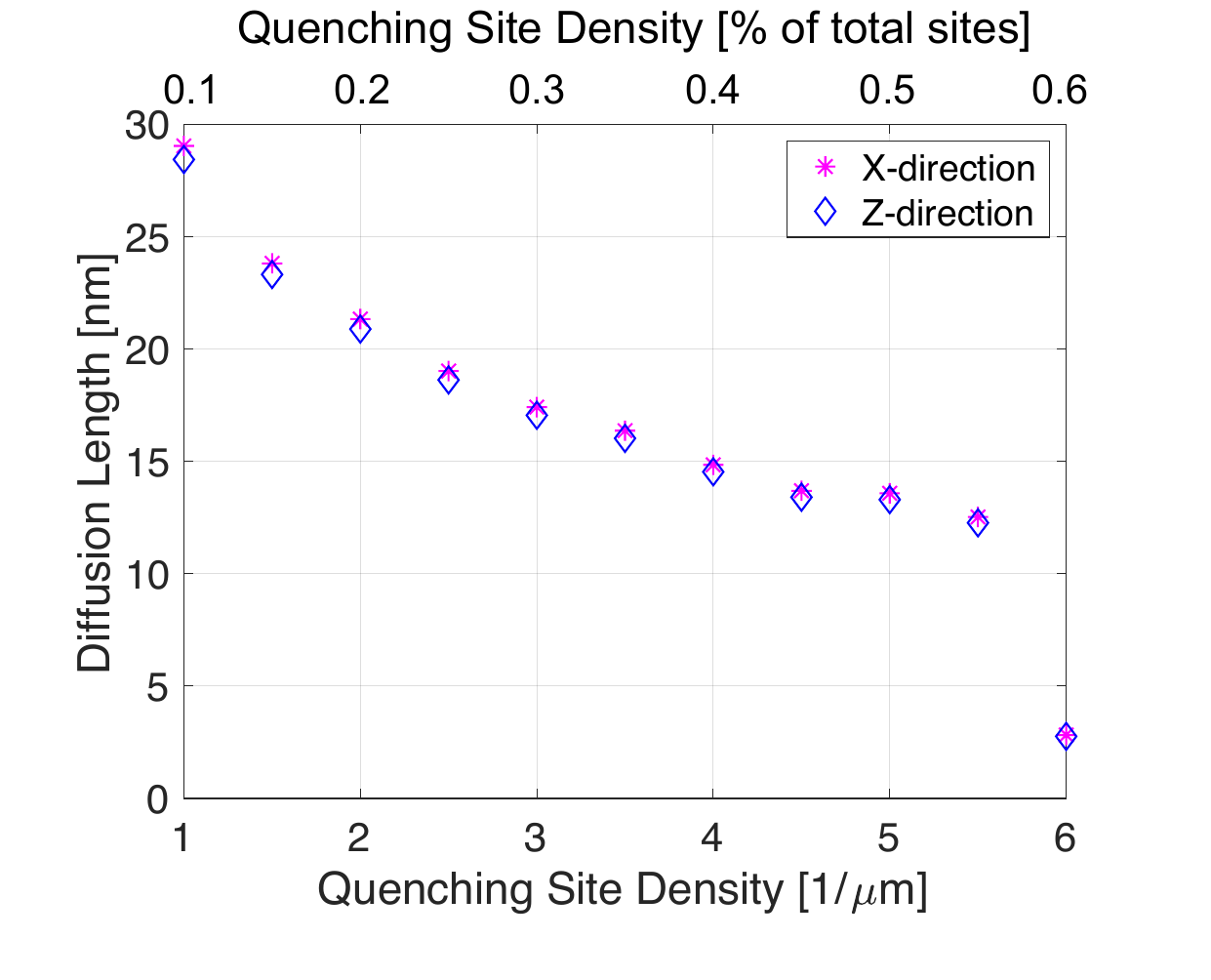}
\caption{Diffusion length for the in-plane ($x$ and $z$) directions vs. quenching-site density in absolute units (bottom axis) and as a percentage (top axis). As expected, diffusion lengths in these directions are nearly identical, and substantially larger (three times) than cross-plane diffusion lengths.}\label{Fig13}
\end{center}
\end{figure}

\subsection{Quenching-Site Density}
Next, we consider the effect of quenching sites ~\cite{Wang_2017_QuenchingSites} on cross-plane diffusion. The effect of quenching sites was described in Sec. \ref{sec:mesh_generation}, but the essence of their inclusion is that excitons within range of a quenching site have all transfer rates set to zero and become stuck for the remainder of the simulation. We randomly select scattering sites to become quenching sites and tune the percentage of quenching sites for a single chirality, (4,2), and morphology, bundled random. Because of our reinjecting boundary condition, all excitons will eventually become quenched in a sufficiently long simulation. Therefore, it does not make sense to calculate the diffusion constant, since its definition as the long-time limit implies it will always be zero when all excitons become quenched. Instead, we calculate the cross-plane diffusion length as a function of quenching-site density as
\begin{equation}
    L_d = \sqrt{D\tau}.
\end{equation}
Here, $L_d$ is the diffusion length, $\tau$ is the average exciton lifetime obtained from a simulation with quenching sites, and $D$ is the diffusion constant calculated for an identical film with no quenching sites. Since there is a cross-plane versus in-plane diffusion anisotropy, we calculate a cross-plane and in-plane diffusion length using the corresponding diagonal elements of the diffusion tensor ($D_{yy}$ for cross-plane and $D_{xx}\approx D_{zz}$ for in-plane diffusion). To compute the exciton lifetime, we determine at what point each exciton stops moving for the duration of the simulation, then average over the ensemble of excitons. In Fig. \ref{Fig12}, we see some interesting behavior. As expected, the diffusion length and lifetime decay with increased quenching-site density. However, we see some critical density at which the plots for both diffusion length and lifetime change curvature and drop sharply. This is because at this density, the exciton lifetime drops to below our simulation time step, indicating that many excitons are quenched before the simulation saves any position data. For comparison, in Fig. \ref{Fig13}, we show the diffusion length and lifetime for the in-plane directions. As expected, the diffusion lengths in both in-plane ($x$ and $z$) directions are very close to one another, as well as considerably larger than the cross-plane diffusion length.

\section{Conclusion}\label{sec:conclusion}

We presented the numerical tool DECaNT that simulates the diffusion of excitons in carbon nanotube thin films. We included resonance energy transfer processes that come from a microscopic theory and capture the effects of orientation, chirality, intertube distance, and environmental effects on exciton transfer rates. The tool uses Bullet Physics, a C++ library for simulating collisions, to create a 3D CNT mesh, which is then fed into our exciton ensemble Monte Carlo tool that tracks exciton position as these quasiparticles diffuse through the film. Based on the position--position correlation function, we calculated the diffusion tensor components and diffusion lengths. This tool is the first of its kind and allows for a new perspective on the behaviour of excitons in CNT films because it can help isolate the effects of parameters that are hard to individually control in experiment. In particular, we studied the impact of film morphology (alignment and bundling), chirality, intertube spacing, and defect density on cross-plane diffusion.

We found that films of aligned CNTs generally have higher cross-plane diffusion than films of random CNTs, because transfer rates are higher between aligned than between misoriented tubes. However, the impact of morphology is strongly dependent on the CNT size. The smallest CNTs, (4,2), have cross-plane diffusion properties that strongly depended on film morphology, while for all other chiralities in the simulation [(6,1),(6,5), and (8,7)], morphology has a much smaller effect. As CNT size increases, the center-to-center distance between scattering sites in different CNTs also increases, and this increase in distance between scattering sites on different tubes overshadows the effect of relative orientation on the transfer rates between CNTs.

We found that adding intertube spacing always reduces cross-plane diffusion, but the rate at which diffusion decreases with distance depends on morphology. The fastest decrease is observed in aligned films, followed by randomly oriented and bundled films.

To study the effect of defect density (``quenching sites'') on the transport properties of CNT films, we calculated the exciton lifetime and diffusion length. We found that in-plane diffusion is isotropic and the associated length is about three times larger than the cross-plane diffusion length. As expected, the diffusion length decreases with increasing quenching-site density.

In short, DECaNT\cite{DECaNT} is a new open-source tool that offers the ability to isolate and study a number of film properties such as morphology, intertube spacing, chirality, and defect density on exciton transport in CNT films. We have presented the effects of key parameters on simulated diffusion. In order to quantitatively compare simulations to measured values, several parameters would generally have to be adjusted at once in DECaNT. Still, some qualitative comparisons can already be made. For example, in Ref. [\onlinecite{Wang_2017_QuenchingSites}], the authors showed that quenching sites strongly affect exciton diffusion length and lifetime, and at quenching-site densities comparable to those in our simulations. Another example comes from Ref. [\onlinecite{flach_2020_improved_exciton_liftime}] for films of (6,5) CNTs, where exciton lifetimes of 4--5 ps and diffusion lengths of 5--10 nm were reported, which would give diffusion constants of about $2\times 10^{-5}\, \mathrm{\mathrm{m^2/s}}$. Our simulated value for the same system, pristine (no quenching sites) and bundled random and with a very small intertube spacing, would be about $9\times 10^{-3}\, \mathrm{m^2/s}$ (see Fig. \ref{Fig10}). The discrepancy is not surprising given our idealized simulations with no added intertube spacing or quenching sites. The combined effects of added intertube spacing (Fig. \ref{Fig11}) and quenching sites (Fig. \ref{Fig12}) would easily lower the diffusion constant to experimental values. A dielectric environment surrounding the tubes would further impede exciton transfer in the simulation.

\section*{Acknowledgement}
The authors thank M. S. Arnold for useful discussions. This work was funded by the U.S. Department of Energy Office of Science (Basic Energy Sciences, Division of Materials
Sciences and Engineering, Physical Behavior of Materials Program) under Award No. DE-SC0008712. Preliminary work was supported in part by the Splinter Professorship (IK) and the NSF UW MRSEC. YCL acknowledges the UW Hilldale Undergraduate Research Fellowship. The simulation work was performed using the compute resources and assistance of the UW-Madison Center For High Throughput Computing (CHTC).


\begin{thebibliography}{47}%
\makeatletter
\providecommand \@ifxundefined [1]{%
 \@ifx{#1\undefined}
}%
\providecommand \@ifnum [1]{%
 \ifnum #1\expandafter \@firstoftwo
 \else \expandafter \@secondoftwo
 \fi
}%
\providecommand \@ifx [1]{%
 \ifx #1\expandafter \@firstoftwo
 \else \expandafter \@secondoftwo
 \fi
}%
\providecommand \natexlab [1]{#1}%
\providecommand \enquote  [1]{``#1''}%
\providecommand \bibnamefont  [1]{#1}%
\providecommand \bibfnamefont [1]{#1}%
\providecommand \citenamefont [1]{#1}%
\providecommand \href@noop [0]{\@secondoftwo}%
\providecommand \href [0]{\begingroup \@sanitize@url \@href}%
\providecommand \@href[1]{\@@startlink{#1}\@@href}%
\providecommand \@@href[1]{\endgroup#1\@@endlink}%
\providecommand \@sanitize@url [0]{\catcode `\\12\catcode `\$12\catcode
  `\&12\catcode `\#12\catcode `\^12\catcode `\_12\catcode `\%12\relax}%
\providecommand \@@startlink[1]{}%
\providecommand \@@endlink[0]{}%
\providecommand \url  [0]{\begingroup\@sanitize@url \@url }%
\providecommand \@url [1]{\endgroup\@href {#1}{\urlprefix }}%
\providecommand \urlprefix  [0]{URL }%
\providecommand \Eprint [0]{\href }%
\providecommand \doibase [0]{https://doi.org/}%
\providecommand \selectlanguage [0]{\@gobble}%
\providecommand \bibinfo  [0]{\@secondoftwo}%
\providecommand \bibfield  [0]{\@secondoftwo}%
\providecommand \translation [1]{[#1]}%
\providecommand \BibitemOpen [0]{}%
\providecommand \bibitemStop [0]{}%
\providecommand \bibitemNoStop [0]{.\EOS\space}%
\providecommand \EOS [0]{\spacefactor3000\relax}%
\providecommand \BibitemShut  [1]{\csname bibitem#1\endcsname}%
\let\auto@bib@innerbib\@empty
\bibitem [{\citenamefont {Jariwala}\ \emph {et~al.}(2013)\citenamefont
  {Jariwala}, \citenamefont {Sangwan}, \citenamefont {Lauhon}, \citenamefont
  {Marks},\ and\ \citenamefont {Hersam}}]{Jariwala_2013_cnt_applications_rev}%
  \BibitemOpen
  \bibfield  {author} {\bibinfo {author} {\bibfnamefont {D.}~\bibnamefont
  {Jariwala}}, \bibinfo {author} {\bibfnamefont {V.~K.}\ \bibnamefont
  {Sangwan}}, \bibinfo {author} {\bibfnamefont {L.~J.}\ \bibnamefont {Lauhon}},
  \bibinfo {author} {\bibfnamefont {T.~J.}\ \bibnamefont {Marks}},\ and\
  \bibinfo {author} {\bibfnamefont {M.~C.}\ \bibnamefont {Hersam}},\
  }\href@noop {} {\bibfield  {journal} {\bibinfo  {journal} {Chem. Soc. Rev.}\
  }\textbf {\bibinfo {volume} {42}},\ \bibinfo {pages} {2824--2860} (\bibinfo
  {year} {2013})}\BibitemShut {NoStop}%
\bibitem [{\citenamefont {Arnold}\ \emph {et~al.}(2013)\citenamefont {Arnold},
  \citenamefont {Blackburn}, \citenamefont {Crochet}, \citenamefont
  {S.~K.~Doorn}, \citenamefont {Mohite},\ and\ \citenamefont
  {Telg}}]{Arnold_2013_cnt_review}%
  \BibitemOpen
  \bibfield  {author} {\bibinfo {author} {\bibfnamefont {M.~S.}\ \bibnamefont
  {Arnold}}, \bibinfo {author} {\bibfnamefont {J.~L.}\ \bibnamefont
  {Blackburn}}, \bibinfo {author} {\bibfnamefont {J.~J.}\ \bibnamefont
  {Crochet}}, \bibinfo {author} {\bibfnamefont {J.~G.~D.}\ \bibnamefont
  {S.~K.~Doorn}}, \bibinfo {author} {\bibfnamefont {A.}~\bibnamefont
  {Mohite}},\ and\ \bibinfo {author} {\bibfnamefont {H.}~\bibnamefont {Telg}},\
  }\href@noop {} {\bibfield  {journal} {\bibinfo  {journal} {Phys. Chem. Chem.
  Phys.}\ }\textbf {\bibinfo {volume} {15}},\ \bibinfo {pages} {14896--14918}
  (\bibinfo {year} {2013})}\BibitemShut {NoStop}%
\bibitem [{\citenamefont {Laird}\ \emph {et~al.}(2015)\citenamefont {Laird},
  \citenamefont {Kuemmeth}, \citenamefont {Steele}, \citenamefont
  {Grove-Rasmussen}, \citenamefont {J.~Nygard},\ and\ \citenamefont
  {Kouwenhoven}}]{Laird_review_2015}%
  \BibitemOpen
  \bibfield  {author} {\bibinfo {author} {\bibfnamefont {E.~A.}\ \bibnamefont
  {Laird}}, \bibinfo {author} {\bibfnamefont {F.}~\bibnamefont {Kuemmeth}},
  \bibinfo {author} {\bibfnamefont {G.~A.}\ \bibnamefont {Steele}}, \bibinfo
  {author} {\bibfnamefont {K.}~\bibnamefont {Grove-Rasmussen}}, \bibinfo
  {author} {\bibfnamefont {K.~F.}\ \bibnamefont {J.~Nygard}},\ and\ \bibinfo
  {author} {\bibfnamefont {L.~P.}\ \bibnamefont {Kouwenhoven}},\ }\href@noop {}
  {\bibfield  {journal} {\bibinfo  {journal} {Rev. Mod. Phys.}\ }\textbf
  {\bibinfo {volume} {87}},\ \bibinfo {pages} {703} (\bibinfo {year}
  {2015})}\BibitemShut {NoStop}%
\bibitem [{\citenamefont {Bindi}\ \emph {et~al.}(2011)\citenamefont {Bindi},
  \citenamefont {Wu}, \citenamefont {Prehn},\ and\ \citenamefont
  {Arnold}}]{Arnold_Films_Harvesting_2011}%
  \BibitemOpen
  \bibfield  {author} {\bibinfo {author} {\bibfnamefont {D.~J.}\ \bibnamefont
  {Bindi}}, \bibinfo {author} {\bibfnamefont {M.}~\bibnamefont {Wu}}, \bibinfo
  {author} {\bibfnamefont {F.~C.}\ \bibnamefont {Prehn}},\ and\ \bibinfo
  {author} {\bibfnamefont {M.~S.}\ \bibnamefont {Arnold}},\ }\href@noop {}
  {\bibfield  {journal} {\bibinfo  {journal} {Nano Lett.}\ }\textbf {\bibinfo
  {volume} {11}},\ \bibinfo {pages} {455--460} (\bibinfo {year}
  {2011})}\BibitemShut {NoStop}%
\bibitem [{\citenamefont {Jeon}\ \emph {et~al.}(2015)\citenamefont {Jeon},
  \citenamefont {Cui}, \citenamefont {Chiba}, \citenamefont {Anisimov},
  \citenamefont {Nasibulin}, \citenamefont {Kauppinen}, \citenamefont
  {Maruyama},\ and\ \citenamefont
  {Matsuo}}]{Jeon_cnt_transparent_electrode_2015}%
  \BibitemOpen
  \bibfield  {author} {\bibinfo {author} {\bibfnamefont {I.}~\bibnamefont
  {Jeon}}, \bibinfo {author} {\bibfnamefont {K.}~\bibnamefont {Cui}}, \bibinfo
  {author} {\bibfnamefont {T.}~\bibnamefont {Chiba}}, \bibinfo {author}
  {\bibfnamefont {A.}~\bibnamefont {Anisimov}}, \bibinfo {author}
  {\bibfnamefont {A.~G.}\ \bibnamefont {Nasibulin}}, \bibinfo {author}
  {\bibfnamefont {E.~I.}\ \bibnamefont {Kauppinen}}, \bibinfo {author}
  {\bibfnamefont {S.}~\bibnamefont {Maruyama}},\ and\ \bibinfo {author}
  {\bibfnamefont {Y.}~\bibnamefont {Matsuo}},\ }\href@noop {} {\bibfield
  {journal} {\bibinfo  {journal} {J. Am. Chem. Soc.}\ }\textbf {\bibinfo
  {volume} {137}},\ \bibinfo {pages} {7982--7985} (\bibinfo {year}
  {2015})}\BibitemShut {NoStop}%
\bibitem [{\citenamefont {Jain}\ \emph {et~al.}(2012)\citenamefont {Jain},
  \citenamefont {Howden}, \citenamefont {Tvrdy}, \citenamefont {Shimizu},
  \citenamefont {Hilmer}, \citenamefont {McNicholas}, \citenamefont {Gleason},\
  and\ \citenamefont {Strano}}]{near_IR_photovoltaics}%
  \BibitemOpen
  \bibfield  {author} {\bibinfo {author} {\bibfnamefont {M.~R.}\ \bibnamefont
  {Jain}}, \bibinfo {author} {\bibfnamefont {R.}~\bibnamefont {Howden}},
  \bibinfo {author} {\bibfnamefont {K.}~\bibnamefont {Tvrdy}}, \bibinfo
  {author} {\bibfnamefont {S.}~\bibnamefont {Shimizu}}, \bibinfo {author}
  {\bibfnamefont {J.~A.}\ \bibnamefont {Hilmer}}, \bibinfo {author}
  {\bibfnamefont {T.~P.}\ \bibnamefont {McNicholas}}, \bibinfo {author}
  {\bibfnamefont {K.~K.}\ \bibnamefont {Gleason}},\ and\ \bibinfo {author}
  {\bibfnamefont {M.~S.}\ \bibnamefont {Strano}},\ }\href@noop {} {\bibfield
  {journal} {\bibinfo  {journal} {Adv. Mater.}\ }\textbf {\bibinfo {volume}
  {24}},\ \bibinfo {pages} {4436--4439} (\bibinfo {year} {2012})}\BibitemShut
  {NoStop}%
\bibitem [{\citenamefont {Flach}\ \emph {et~al.}(2020)\citenamefont {Flach},
  \citenamefont {Wang}, \citenamefont {Arnold},\ and\ \citenamefont
  {Zanni}}]{flach_2020_improved_exciton_liftime}%
  \BibitemOpen
  \bibfield  {author} {\bibinfo {author} {\bibfnamefont {J.~T.}\ \bibnamefont
  {Flach}}, \bibinfo {author} {\bibfnamefont {J.}~\bibnamefont {Wang}},
  \bibinfo {author} {\bibfnamefont {M.~S.}\ \bibnamefont {Arnold}},\ and\
  \bibinfo {author} {\bibfnamefont {M.~T.}\ \bibnamefont {Zanni}},\ }\href@noop
  {} {\bibfield  {journal} {\bibinfo  {journal} {Journal of Physical Chemistry
  Letters}\ }\textbf {\bibinfo {volume} {11}},\ \bibinfo {pages} {6016 -- 6024}
  (\bibinfo {year} {2020})}\BibitemShut {NoStop}%
\bibitem [{\citenamefont {Ruzicka.}\ \emph {et~al.}(2012)\citenamefont
  {Ruzicka.}, \citenamefont {Wang}, \citenamefont {Lohrman}, \citenamefont
  {Ren},\ and\ \citenamefont {Zhao}}]{Ruzicka_2012_exciton_diffusion_constant}%
  \BibitemOpen
  \bibfield  {author} {\bibinfo {author} {\bibfnamefont {B.~A.}\ \bibnamefont
  {Ruzicka.}}, \bibinfo {author} {\bibfnamefont {R.}~\bibnamefont {Wang}},
  \bibinfo {author} {\bibfnamefont {J.}~\bibnamefont {Lohrman}}, \bibinfo
  {author} {\bibfnamefont {S.}~\bibnamefont {Ren}},\ and\ \bibinfo {author}
  {\bibfnamefont {H.}~\bibnamefont {Zhao}},\ }\href@noop {} {\bibfield
  {journal} {\bibinfo  {journal} {Phys. Rev. B}\ }\textbf {\bibinfo {volume}
  {86}},\ \bibinfo {pages} {205417} (\bibinfo {year} {2012})}\BibitemShut
  {NoStop}%
\bibitem [{\citenamefont {Shea}\ and\ \citenamefont
  {Arnold}(2013{\natexlab{a}})}]{Shea_2013_exciton_diff_length}%
  \BibitemOpen
  \bibfield  {author} {\bibinfo {author} {\bibfnamefont {M.~J.}\ \bibnamefont
  {Shea}}\ and\ \bibinfo {author} {\bibfnamefont {M.~S.}\ \bibnamefont
  {Arnold}},\ }\href@noop {} {\bibfield  {journal} {\bibinfo  {journal}
  {Applied Physics Letters}\ }\textbf {\bibinfo {volume} {102}},\ \bibinfo
  {pages} {243101} (\bibinfo {year} {2013}{\natexlab{a}})}\BibitemShut
  {NoStop}%
\bibitem [{\citenamefont {Dowgiallo}\ \emph {et~al.}(2016)\citenamefont
  {Dowgiallo}, \citenamefont {Mistry}, \citenamefont {Johnson}, \citenamefont
  {Reid},\ and\ \citenamefont {Blackburn}}]{CNT_c60_interface_Dowgiallo_2016}%
  \BibitemOpen
  \bibfield  {author} {\bibinfo {author} {\bibfnamefont {A.}~\bibnamefont
  {Dowgiallo}}, \bibinfo {author} {\bibfnamefont {K.~S.}\ \bibnamefont
  {Mistry}}, \bibinfo {author} {\bibfnamefont {J.~C.}\ \bibnamefont {Johnson}},
  \bibinfo {author} {\bibfnamefont {O.~G.}\ \bibnamefont {Reid}},\ and\
  \bibinfo {author} {\bibfnamefont {J.~L.}\ \bibnamefont {Blackburn}},\
  }\href@noop {} {\bibfield  {journal} {\bibinfo  {journal} {J. Phys. Chem.
  Lett.}\ }\textbf {\bibinfo {volume} {7}},\ \bibinfo {pages} {1794--1799}
  (\bibinfo {year} {2016})}\BibitemShut {NoStop}%
\bibitem [{\citenamefont {Postupna}, \citenamefont {Jaeger},\ and\
  \citenamefont {Prezhdo}(2014)}]{Postupna_2014_JPCL}%
  \BibitemOpen
  \bibfield  {author} {\bibinfo {author} {\bibfnamefont {O.}~\bibnamefont
  {Postupna}}, \bibinfo {author} {\bibfnamefont {H.~M.}\ \bibnamefont
  {Jaeger}},\ and\ \bibinfo {author} {\bibfnamefont {O.~V.}\ \bibnamefont
  {Prezhdo}},\ }\bibfield  {title} {\enquote {\bibinfo {title} {Photoinduced
  dynamics in carbon nanotube aggregates steered by dark excitons},}\ }\href
  {https://doi.org/10.1021/jz502052b} {\bibfield  {journal} {\bibinfo
  {journal} {The Journal of Physical Chemistry Letters}\ }\textbf {\bibinfo
  {volume} {5}},\ \bibinfo {pages} {3872--3877} (\bibinfo {year} {2014})},\
  \bibinfo {note} {pMID: 26278762},\ \Eprint
  {https://arxiv.org/abs/https://doi.org/10.1021/jz502052b}
  {https://doi.org/10.1021/jz502052b} \BibitemShut {NoStop}%
\bibitem [{\citenamefont {Wong}\ \emph {et~al.}(2009)\citenamefont {Wong},
  \citenamefont {Curutchet}, \citenamefont {Tretiak},\ and\ \citenamefont
  {Scholes}}]{Wong_2009_JCP}%
  \BibitemOpen
  \bibfield  {author} {\bibinfo {author} {\bibfnamefont {C.~Y.}\ \bibnamefont
  {Wong}}, \bibinfo {author} {\bibfnamefont {C.}~\bibnamefont {Curutchet}},
  \bibinfo {author} {\bibfnamefont {S.}~\bibnamefont {Tretiak}},\ and\ \bibinfo
  {author} {\bibfnamefont {G.~D.}\ \bibnamefont {Scholes}},\ }\bibfield
  {title} {\enquote {\bibinfo {title} {Ideal dipole approximation fails to
  predict electronic coupling and energy transfer between semiconducting
  single-wall carbon nanotubes},}\ }\href {https://doi.org/10.1063/1.3088846}
  {\bibfield  {journal} {\bibinfo  {journal} {The Journal of Chemical Physics}\
  }\textbf {\bibinfo {volume} {130}},\ \bibinfo {pages} {081104} (\bibinfo
  {year} {2009})},\ \Eprint
  {https://arxiv.org/abs/https://doi.org/10.1063/1.3088846}
  {https://doi.org/10.1063/1.3088846} \BibitemShut {NoStop}%
\bibitem [{\citenamefont {Jones}\ and\ \citenamefont
  {Bradshaw}(2019)}]{Jones_2019_RET_review}%
  \BibitemOpen
  \bibfield  {author} {\bibinfo {author} {\bibfnamefont {G.~A.}\ \bibnamefont
  {Jones}}\ and\ \bibinfo {author} {\bibfnamefont {D.~S.}\ \bibnamefont
  {Bradshaw}},\ }\bibfield  {title} {\enquote {\bibinfo {title} {Resonance
  energy transfer: From fundamental theory to recent applications},}\ }\href
  {https://doi.org/10.3389/fphy.2019.00100} {\bibfield  {journal} {\bibinfo
  {journal} {Frontiers in Physics}\ }\textbf {\bibinfo {volume} {7}},\ \bibinfo
  {pages} {100} (\bibinfo {year} {2019})}\BibitemShut {NoStop}%
\bibitem [{\citenamefont {Davoody}\ \emph {et~al.}(2016)\citenamefont
  {Davoody}, \citenamefont {Karimi}, \citenamefont {Arnold},\ and\
  \citenamefont {Knezevic}}]{Davoody_exciton_transfer}%
  \BibitemOpen
  \bibfield  {author} {\bibinfo {author} {\bibfnamefont {A.~H.}\ \bibnamefont
  {Davoody}}, \bibinfo {author} {\bibfnamefont {F.}~\bibnamefont {Karimi}},
  \bibinfo {author} {\bibfnamefont {M.~S.}\ \bibnamefont {Arnold}},\ and\
  \bibinfo {author} {\bibfnamefont {I.}~\bibnamefont {Knezevic}},\ }\href@noop
  {} {\bibfield  {journal} {\bibinfo  {journal} {J. Phys. Chem. C}\ }\textbf
  {\bibinfo {volume} {120}},\ \bibinfo {pages} {16354--16366} (\bibinfo {year}
  {2016})}\BibitemShut {NoStop}%
\bibitem [{\citenamefont {Davoody}\ \emph {et~al.}(2017)\citenamefont
  {Davoody}, \citenamefont {Karimi}, \citenamefont {Arnold},\ and\
  \citenamefont {Knezevic}}]{Davoody_exciton_phonon_transfer}%
  \BibitemOpen
  \bibfield  {author} {\bibinfo {author} {\bibfnamefont {A.~H.}\ \bibnamefont
  {Davoody}}, \bibinfo {author} {\bibfnamefont {F.}~\bibnamefont {Karimi}},
  \bibinfo {author} {\bibfnamefont {M.~S.}\ \bibnamefont {Arnold}},\ and\
  \bibinfo {author} {\bibfnamefont {I.}~\bibnamefont {Knezevic}},\ }\href@noop
  {} {\bibfield  {journal} {\bibinfo  {journal} {J. Phys. Chem. C}\ }\textbf
  {\bibinfo {volume} {121}},\ \bibinfo {pages} {13084--13091} (\bibinfo {year}
  {2017})}\BibitemShut {NoStop}%
\bibitem [{\citenamefont {Li}\ \emph {et~al.}(2020)\citenamefont {Li},
  \citenamefont {Davoody}, \citenamefont {Belling}, \citenamefont {Gabourie},\
  and\ \citenamefont {Knezevic}}]{DECaNT}%
  \BibitemOpen
  \bibfield  {author} {\bibinfo {author} {\bibfnamefont {Y.~C.}\ \bibnamefont
  {Li}}, \bibinfo {author} {\bibfnamefont {A.~H.}\ \bibnamefont {Davoody}},
  \bibinfo {author} {\bibfnamefont {S.~W.}\ \bibnamefont {Belling}}, \bibinfo
  {author} {\bibfnamefont {A.~J.}\ \bibnamefont {Gabourie}},\ and\ \bibinfo
  {author} {\bibfnamefont {I.}~\bibnamefont {Knezevic}},\ }\href@noop {}
  {\enquote {\bibinfo {title} {{DECaNT}: {D}iffusion of {E}xcitons in {C}arbon
  {N}ano{T}ubes},}\ }\bibinfo {howpublished} {https://github.com/li779/DECaNT}
  (\bibinfo {year} {2020})\BibitemShut {NoStop}%
\bibitem [{\citenamefont {Coumans}(2020)}]{bullet_physics}%
  \BibitemOpen
  \bibfield  {author} {\bibinfo {author} {\bibfnamefont {E.}~\bibnamefont
  {Coumans}},\ }\href@noop {} {\enquote {\bibinfo {title} {Bullet {P}hysics
  library},}\ }\bibinfo {howpublished} {http://bulletphysics.org} (\bibinfo
  {year} {2020})\BibitemShut {NoStop}%
\bibitem [{\citenamefont {Avouris}, \citenamefont {Chen},\ and\ \citenamefont
  {Perebeinos}(2007)}]{Avouris_CNT_Electronics}%
  \BibitemOpen
  \bibfield  {author} {\bibinfo {author} {\bibfnamefont {P.}~\bibnamefont
  {Avouris}}, \bibinfo {author} {\bibfnamefont {Z.}~\bibnamefont {Chen}},\ and\
  \bibinfo {author} {\bibfnamefont {V.}~\bibnamefont {Perebeinos}},\
  }\href@noop {} {\bibfield  {journal} {\bibinfo  {journal} {Nat. Nanotech.}\
  }\textbf {\bibinfo {volume} {2}},\ \bibinfo {pages} {605--615} (\bibinfo
  {year} {2007})}\BibitemShut {NoStop}%
\bibitem [{\citenamefont {A.~L.~Fetter}(1971)}]{many_body_quantum_Fetter_1971}%
  \BibitemOpen
  \bibfield  {author} {\bibinfo {author} {\bibfnamefont {J.~D.~W.}\
  \bibnamefont {A.~L.~Fetter}},\ }\href@noop {} {\emph {\bibinfo {title}
  {Quantum Theory of Many Particle Systems}}}\ (\bibinfo  {publisher}
  {McGraw-Hill},\ \bibinfo {year} {1971})\BibitemShut {NoStop}%
\bibitem [{\citenamefont {Reich}\ \emph {et~al.}(2002)\citenamefont {Reich},
  \citenamefont {Maultzsch}, \citenamefont {Thomsen},\ and\ \citenamefont
  {Ordejon}}]{graphene_tightbinding_Reich_2002}%
  \BibitemOpen
  \bibfield  {author} {\bibinfo {author} {\bibfnamefont {S.}~\bibnamefont
  {Reich}}, \bibinfo {author} {\bibfnamefont {J.}~\bibnamefont {Maultzsch}},
  \bibinfo {author} {\bibfnamefont {C.}~\bibnamefont {Thomsen}},\ and\ \bibinfo
  {author} {\bibfnamefont {P.}~\bibnamefont {Ordejon}},\ }\href@noop {}
  {\bibfield  {journal} {\bibinfo  {journal} {Phys. Rev. B}\ }\textbf {\bibinfo
  {volume} {66}},\ \bibinfo {pages} {035412} (\bibinfo {year}
  {2002})}\BibitemShut {NoStop}%
\bibitem [{\citenamefont {Saito}, \citenamefont {Dresselhaus},\ and\
  \citenamefont {Dresselhaus}(1998)}]{Dresselhaus_CNT_Properties}%
  \BibitemOpen
  \bibfield  {author} {\bibinfo {author} {\bibfnamefont {R.}~\bibnamefont
  {Saito}}, \bibinfo {author} {\bibfnamefont {G.}~\bibnamefont {Dresselhaus}},\
  and\ \bibinfo {author} {\bibfnamefont {M.~S.}\ \bibnamefont {Dresselhaus}},\
  }\href@noop {} {\emph {\bibinfo {title} {Physical Properties of Carbon
  Nanotubes}}}\ (\bibinfo  {publisher} {World Scientific},\ \bibinfo {year}
  {1998})\BibitemShut {NoStop}%
\bibitem [{\citenamefont {Strinati}(1988)}]{greens_functions}%
  \BibitemOpen
  \bibfield  {author} {\bibinfo {author} {\bibfnamefont {G.}~\bibnamefont
  {Strinati}},\ }\href@noop {} {\bibfield  {journal} {\bibinfo  {journal} {La
  Rivista del Nuovo Cimento}\ }\textbf {\bibinfo {volume} {11}},\ \bibinfo
  {pages} {1--86} (\bibinfo {year} {1988})}\BibitemShut {NoStop}%
\bibitem [{\citenamefont {Barros}\ \emph
  {et~al.}(2006{\natexlab{a}})\citenamefont {Barros}, \citenamefont {Jorio},
  \citenamefont {Samsonidze}, \citenamefont {Capaz}, \citenamefont {Filho},
  \citenamefont {Filho}, \citenamefont {Dresselhaus},\ and\ \citenamefont
  {Dresselhaus}}]{Dresselhaus_symmetry_2006_phys_rep}%
  \BibitemOpen
  \bibfield  {author} {\bibinfo {author} {\bibfnamefont {E.~B.}\ \bibnamefont
  {Barros}}, \bibinfo {author} {\bibfnamefont {A.}~\bibnamefont {Jorio}},
  \bibinfo {author} {\bibfnamefont {G.~G.}\ \bibnamefont {Samsonidze}},
  \bibinfo {author} {\bibfnamefont {R.~B.}\ \bibnamefont {Capaz}}, \bibinfo
  {author} {\bibfnamefont {A.~G.~S.}\ \bibnamefont {Filho}}, \bibinfo {author}
  {\bibfnamefont {J.~M.}\ \bibnamefont {Filho}}, \bibinfo {author}
  {\bibfnamefont {G.}~\bibnamefont {Dresselhaus}},\ and\ \bibinfo {author}
  {\bibfnamefont {M.~S.}\ \bibnamefont {Dresselhaus}},\ }\href@noop {}
  {\bibfield  {journal} {\bibinfo  {journal} {Phys. Rep.}\ }\textbf {\bibinfo
  {volume} {431}},\ \bibinfo {pages} {261--302} (\bibinfo {year}
  {2006}{\natexlab{a}})}\BibitemShut {NoStop}%
\bibitem [{\citenamefont {Barros}\ \emph
  {et~al.}(2006{\natexlab{b}})\citenamefont {Barros}, \citenamefont {Capaz},
  \citenamefont {Jorio}, \citenamefont {Samsonidze}, \citenamefont {Filho},
  \citenamefont {Ismail-Beigi}, \citenamefont {Spataru}, \citenamefont {Louie},
  \citenamefont {Dresselhaus},\ and\ \citenamefont
  {Dresselhaus}}]{Dresselhaus_symmetry_2006}%
  \BibitemOpen
  \bibfield  {author} {\bibinfo {author} {\bibfnamefont {E.~B.}\ \bibnamefont
  {Barros}}, \bibinfo {author} {\bibfnamefont {R.~B.}\ \bibnamefont {Capaz}},
  \bibinfo {author} {\bibfnamefont {A.}~\bibnamefont {Jorio}}, \bibinfo
  {author} {\bibfnamefont {G.~G.}\ \bibnamefont {Samsonidze}}, \bibinfo
  {author} {\bibfnamefont {A.~G.~S.}\ \bibnamefont {Filho}}, \bibinfo {author}
  {\bibfnamefont {S.}~\bibnamefont {Ismail-Beigi}}, \bibinfo {author}
  {\bibfnamefont {C.~D.}\ \bibnamefont {Spataru}}, \bibinfo {author}
  {\bibfnamefont {S.~G.}\ \bibnamefont {Louie}}, \bibinfo {author}
  {\bibfnamefont {G.}~\bibnamefont {Dresselhaus}},\ and\ \bibinfo {author}
  {\bibfnamefont {M.~S.}\ \bibnamefont {Dresselhaus}},\ }\href@noop {}
  {\bibfield  {journal} {\bibinfo  {journal} {Phys. Rev. B}\ }\textbf {\bibinfo
  {volume} {73}},\ \bibinfo {pages} {241406} (\bibinfo {year}
  {2006}{\natexlab{b}})}\BibitemShut {NoStop}%
\bibitem [{\citenamefont {Jishi}, \citenamefont {Dresselhaus},\ and\
  \citenamefont {Dresselhaus}(1993)}]{Dresselhaus_symmetry_1993}%
  \BibitemOpen
  \bibfield  {author} {\bibinfo {author} {\bibfnamefont {R.~A.}\ \bibnamefont
  {Jishi}}, \bibinfo {author} {\bibfnamefont {G.}~\bibnamefont {Dresselhaus}},\
  and\ \bibinfo {author} {\bibfnamefont {M.~S.}\ \bibnamefont {Dresselhaus}},\
  }\href@noop {} {\bibfield  {journal} {\bibinfo  {journal} {Phys. Rev. B}\
  }\textbf {\bibinfo {volume} {47}},\ \bibinfo {pages} {16671--16674} (\bibinfo
  {year} {1993})}\BibitemShut {NoStop}%
\bibitem [{\citenamefont {Jiang}\ \emph {et~al.}(2007)\citenamefont {Jiang},
  \citenamefont {Saito}, \citenamefont {Samsonidze}, \citenamefont {Jorio},
  \citenamefont {Chou}, \citenamefont {Dresselhaus},\ and\ \citenamefont
  {Dresselhaus}}]{dresselhaus_CNT_exciton_TB}%
  \BibitemOpen
  \bibfield  {author} {\bibinfo {author} {\bibfnamefont {J.}~\bibnamefont
  {Jiang}}, \bibinfo {author} {\bibfnamefont {R.}~\bibnamefont {Saito}},
  \bibinfo {author} {\bibfnamefont {G.~G.}\ \bibnamefont {Samsonidze}},
  \bibinfo {author} {\bibfnamefont {A.}~\bibnamefont {Jorio}}, \bibinfo
  {author} {\bibfnamefont {S.~G.}\ \bibnamefont {Chou}}, \bibinfo {author}
  {\bibfnamefont {G.}~\bibnamefont {Dresselhaus}},\ and\ \bibinfo {author}
  {\bibfnamefont {M.~S.}\ \bibnamefont {Dresselhaus}},\ }\href@noop {}
  {\bibfield  {journal} {\bibinfo  {journal} {Phys. Rev. B}\ }\textbf {\bibinfo
  {volume} {75}},\ \bibinfo {pages} {035407} (\bibinfo {year}
  {2007})}\BibitemShut {NoStop}%
\bibitem [{\citenamefont {Jinkins}\ \emph
  {et~al.}(2019{\natexlab{a}})\citenamefont {Jinkins}, \citenamefont {Chan},
  \citenamefont {Jacobberger}, \citenamefont {Berson},\ and\ \citenamefont
  {Arnold}}]{alignment_jinkins_2019}%
  \BibitemOpen
  \bibfield  {author} {\bibinfo {author} {\bibfnamefont {K.~R.}\ \bibnamefont
  {Jinkins}}, \bibinfo {author} {\bibfnamefont {J.}~\bibnamefont {Chan}},
  \bibinfo {author} {\bibfnamefont {R.~M.}\ \bibnamefont {Jacobberger}},
  \bibinfo {author} {\bibfnamefont {A.}~\bibnamefont {Berson}},\ and\ \bibinfo
  {author} {\bibfnamefont {M.~S.}\ \bibnamefont {Arnold}},\ }\href@noop {}
  {\bibfield  {journal} {\bibinfo  {journal} {ECS Transactions}\ }\textbf
  {\bibinfo {volume} {93}},\ \bibinfo {pages} {117--120} (\bibinfo {year}
  {2019}{\natexlab{a}})}\BibitemShut {NoStop}%
\bibitem [{\citenamefont {Jinkins}\ \emph {et~al.}(2017)\citenamefont
  {Jinkins}, \citenamefont {Chan}, \citenamefont {Brady}, \citenamefont
  {Gronski}, \citenamefont {Gopalan}, \citenamefont {Evensen}, \citenamefont
  {Berson},\ and\ \citenamefont {Arnold}}]{CNT_alignment_arnold}%
  \BibitemOpen
  \bibfield  {author} {\bibinfo {author} {\bibfnamefont {K.~R.}\ \bibnamefont
  {Jinkins}}, \bibinfo {author} {\bibfnamefont {J.}~\bibnamefont {Chan}},
  \bibinfo {author} {\bibfnamefont {G.~J.}\ \bibnamefont {Brady}}, \bibinfo
  {author} {\bibfnamefont {K.~K.}\ \bibnamefont {Gronski}}, \bibinfo {author}
  {\bibfnamefont {P.}~\bibnamefont {Gopalan}}, \bibinfo {author} {\bibfnamefont
  {H.~T.}\ \bibnamefont {Evensen}}, \bibinfo {author} {\bibfnamefont
  {A.}~\bibnamefont {Berson}},\ and\ \bibinfo {author} {\bibfnamefont {M.~S.}\
  \bibnamefont {Arnold}},\ }\href@noop {} {\bibfield  {journal} {\bibinfo
  {journal} {Langmuir}\ }\textbf {\bibinfo {volume} {33}},\ \bibinfo {pages}
  {13407--13414} (\bibinfo {year} {2017})}\BibitemShut {NoStop}%
\bibitem [{\citenamefont {Wei}\ \emph {et~al.}(2018)\citenamefont {Wei},
  \citenamefont {Tanaka}, \citenamefont {Hirakawa}, \citenamefont {Tsuzuki},
  \citenamefont {Wang}, \citenamefont {Yomogida}, \citenamefont {Hirano},\ and\
  \citenamefont {Kataura}}]{Wei_chiral_enantiomer_separation_2018}%
  \BibitemOpen
  \bibfield  {author} {\bibinfo {author} {\bibfnamefont {X.}~\bibnamefont
  {Wei}}, \bibinfo {author} {\bibfnamefont {T.}~\bibnamefont {Tanaka}},
  \bibinfo {author} {\bibfnamefont {T.}~\bibnamefont {Hirakawa}}, \bibinfo
  {author} {\bibfnamefont {M.}~\bibnamefont {Tsuzuki}}, \bibinfo {author}
  {\bibfnamefont {G.}~\bibnamefont {Wang}}, \bibinfo {author} {\bibfnamefont
  {Y.}~\bibnamefont {Yomogida}}, \bibinfo {author} {\bibfnamefont
  {A.}~\bibnamefont {Hirano}},\ and\ \bibinfo {author} {\bibfnamefont
  {H.}~\bibnamefont {Kataura}},\ }\href@noop {} {\bibfield  {journal} {\bibinfo
   {journal} {Carbon}\ }\textbf {\bibinfo {volume} {132}},\ \bibinfo {pages}
  {1--7} (\bibinfo {year} {2018})}\BibitemShut {NoStop}%
\bibitem [{\citenamefont {Ihara}\ \emph {et~al.}(2011)\citenamefont {Ihara},
  \citenamefont {Endoh}, \citenamefont {Saito},\ and\ \citenamefont
  {Nihey}}]{Ihara_metal_separation_2011}%
  \BibitemOpen
  \bibfield  {author} {\bibinfo {author} {\bibfnamefont {K.}~\bibnamefont
  {Ihara}}, \bibinfo {author} {\bibfnamefont {H.}~\bibnamefont {Endoh}},
  \bibinfo {author} {\bibfnamefont {T.}~\bibnamefont {Saito}},\ and\ \bibinfo
  {author} {\bibfnamefont {F.}~\bibnamefont {Nihey}},\ }\href@noop {}
  {\bibfield  {journal} {\bibinfo  {journal} {J. Phys. Chem. C}\ }\textbf
  {\bibinfo {volume} {115}},\ \bibinfo {pages} {22827--22832} (\bibinfo {year}
  {2011})}\BibitemShut {NoStop}%
\bibitem [{\citenamefont {Sturzl}\ \emph {et~al.}(2009)\citenamefont {Sturzl},
  \citenamefont {Hennrich}, \citenamefont {Lebedkin},\ and\ \citenamefont
  {Kappes}}]{Sturzl_chiral_separation_2009}%
  \BibitemOpen
  \bibfield  {author} {\bibinfo {author} {\bibfnamefont {N.}~\bibnamefont
  {Sturzl}}, \bibinfo {author} {\bibfnamefont {F.}~\bibnamefont {Hennrich}},
  \bibinfo {author} {\bibfnamefont {S.}~\bibnamefont {Lebedkin}},\ and\
  \bibinfo {author} {\bibfnamefont {M.~M.}\ \bibnamefont {Kappes}},\
  }\href@noop {} {\bibfield  {journal} {\bibinfo  {journal} {J. Phys. Chem. C}\
  }\textbf {\bibinfo {volume} {113}},\ \bibinfo {pages} {14628--14632}
  (\bibinfo {year} {2009})}\BibitemShut {NoStop}%
\bibitem [{\citenamefont {Yomogida}\ \emph {et~al.}(2016)\citenamefont
  {Yomogida}, \citenamefont {Tanaka}, \citenamefont {Zhang}, \citenamefont
  {Yudasaka}, \citenamefont {Wei},\ and\ \citenamefont
  {Kataura}}]{Yomogida_chiral_separation_2016}%
  \BibitemOpen
  \bibfield  {author} {\bibinfo {author} {\bibfnamefont {Y.}~\bibnamefont
  {Yomogida}}, \bibinfo {author} {\bibfnamefont {T.}~\bibnamefont {Tanaka}},
  \bibinfo {author} {\bibfnamefont {M.}~\bibnamefont {Zhang}}, \bibinfo
  {author} {\bibfnamefont {M.}~\bibnamefont {Yudasaka}}, \bibinfo {author}
  {\bibfnamefont {X.}~\bibnamefont {Wei}},\ and\ \bibinfo {author}
  {\bibfnamefont {H.}~\bibnamefont {Kataura}},\ }\href@noop {} {\bibfield
  {journal} {\bibinfo  {journal} {Nat. Comm.}\ }\textbf {\bibinfo {volume}
  {7}},\ \bibinfo {pages} {12056} (\bibinfo {year} {2016})}\BibitemShut
  {NoStop}%
\bibitem [{\citenamefont {Graf}\ \emph {et~al.}(2016)\citenamefont {Graf},
  \citenamefont {Zakharko}, \citenamefont {Schießl}, \citenamefont {Backes},
  \citenamefont {Pfohl}, \citenamefont {Flavel},\ and\ \citenamefont
  {Zaumseil}}]{Graf_2016_chiral_separation}%
  \BibitemOpen
  \bibfield  {author} {\bibinfo {author} {\bibfnamefont {A.}~\bibnamefont
  {Graf}}, \bibinfo {author} {\bibfnamefont {Y.}~\bibnamefont {Zakharko}},
  \bibinfo {author} {\bibfnamefont {S.~P.}\ \bibnamefont {Schießl}}, \bibinfo
  {author} {\bibfnamefont {C.}~\bibnamefont {Backes}}, \bibinfo {author}
  {\bibfnamefont {M.}~\bibnamefont {Pfohl}}, \bibinfo {author} {\bibfnamefont
  {B.~S.}\ \bibnamefont {Flavel}},\ and\ \bibinfo {author} {\bibfnamefont
  {J.}~\bibnamefont {Zaumseil}},\ }\href@noop {} {\bibfield  {journal}
  {\bibinfo  {journal} {Carbon}\ }\textbf {\bibinfo {volume} {105}},\ \bibinfo
  {pages} {593 -- 599} (\bibinfo {year} {2016})}\BibitemShut {NoStop}%
\bibitem [{\citenamefont {Mehlenbacher}\ \emph {et~al.}(2016)\citenamefont
  {Mehlenbacher}, \citenamefont {McDonough}, \citenamefont {Kearns},
  \citenamefont {Shea}, \citenamefont {Joo}, \citenamefont {Gopalan},
  \citenamefont {Arnold}, ,\ and\ \citenamefont
  {Zanni}}]{mehlenbacher_2016_cnt_films_spectroscopy}%
  \BibitemOpen
  \bibfield  {author} {\bibinfo {author} {\bibfnamefont {R.~D.}\ \bibnamefont
  {Mehlenbacher}}, \bibinfo {author} {\bibfnamefont {T.~J.}\ \bibnamefont
  {McDonough}}, \bibinfo {author} {\bibfnamefont {N.~M.}\ \bibnamefont
  {Kearns}}, \bibinfo {author} {\bibfnamefont {M.~J.}\ \bibnamefont {Shea}},
  \bibinfo {author} {\bibfnamefont {Y.}~\bibnamefont {Joo}}, \bibinfo {author}
  {\bibfnamefont {P.}~\bibnamefont {Gopalan}}, \bibinfo {author} {\bibfnamefont
  {M.~S.}\ \bibnamefont {Arnold}}, ,\ and\ \bibinfo {author} {\bibfnamefont
  {M.~T.}\ \bibnamefont {Zanni}},\ }\href@noop {} {\bibfield  {journal}
  {\bibinfo  {journal} {J. Phys. Chem. C}\ }\textbf {\bibinfo {volume} {120}},\
  \bibinfo {pages} {17069--17080} (\bibinfo {year} {2016})}\BibitemShut
  {NoStop}%
\bibitem [{\citenamefont {Shea}\ and\ \citenamefont
  {Arnold}(2013{\natexlab{b}})}]{film_thickness_Shea_2013}%
  \BibitemOpen
  \bibfield  {author} {\bibinfo {author} {\bibfnamefont {M.~J.}\ \bibnamefont
  {Shea}}\ and\ \bibinfo {author} {\bibfnamefont {M.~S.}\ \bibnamefont
  {Arnold}},\ }\href@noop {} {\bibfield  {journal} {\bibinfo  {journal} {Appl.
  Phys. Lett.}\ }\textbf {\bibinfo {volume} {102}},\ \bibinfo {pages} {243101}
  (\bibinfo {year} {2013}{\natexlab{b}})}\BibitemShut {NoStop}%
\bibitem [{\citenamefont {Streit}\ \emph {et~al.}(2012)\citenamefont {Streit},
  \citenamefont {Bachilo}, \citenamefont {Naumov}, \citenamefont {Khripin},
  \citenamefont {Zheng},\ and\ \citenamefont
  {Weisman}}]{CNT_length_Streit_2012}%
  \BibitemOpen
  \bibfield  {author} {\bibinfo {author} {\bibfnamefont {J.~K.}\ \bibnamefont
  {Streit}}, \bibinfo {author} {\bibfnamefont {S.~M.}\ \bibnamefont {Bachilo}},
  \bibinfo {author} {\bibfnamefont {A.~V.}\ \bibnamefont {Naumov}}, \bibinfo
  {author} {\bibfnamefont {C.}~\bibnamefont {Khripin}}, \bibinfo {author}
  {\bibfnamefont {M.}~\bibnamefont {Zheng}},\ and\ \bibinfo {author}
  {\bibfnamefont {R.~B.}\ \bibnamefont {Weisman}},\ }\href@noop {} {\bibfield
  {journal} {\bibinfo  {journal} {ACS Nano}\ }\textbf {\bibinfo {volume} {6}},\
  \bibinfo {pages} {8424--8431} (\bibinfo {year} {2012})}\BibitemShut {NoStop}%
\bibitem [{\citenamefont {Cantoro}\ \emph {et~al.}(2006)\citenamefont
  {Cantoro}, \citenamefont {Hoffman}, \citenamefont {Pisana}, \citenamefont
  {Scardaci}, \citenamefont {Parvez}, \citenamefont {Ducati}, \citenamefont
  {Ferrari}, \citenamefont {Blackburn}, \citenamefont {Wang},\ and\
  \citenamefont {Robertson}}]{cantoro_alignment_2006}%
  \BibitemOpen
  \bibfield  {author} {\bibinfo {author} {\bibfnamefont {M.}~\bibnamefont
  {Cantoro}}, \bibinfo {author} {\bibfnamefont {S.}~\bibnamefont {Hoffman}},
  \bibinfo {author} {\bibfnamefont {S.}~\bibnamefont {Pisana}}, \bibinfo
  {author} {\bibfnamefont {V.}~\bibnamefont {Scardaci}}, \bibinfo {author}
  {\bibfnamefont {A.}~\bibnamefont {Parvez}}, \bibinfo {author} {\bibfnamefont
  {C.}~\bibnamefont {Ducati}}, \bibinfo {author} {\bibfnamefont
  {A.}~\bibnamefont {Ferrari}}, \bibinfo {author} {\bibfnamefont {A.~M.}\
  \bibnamefont {Blackburn}}, \bibinfo {author} {\bibfnamefont {K.}~\bibnamefont
  {Wang}},\ and\ \bibinfo {author} {\bibfnamefont {J.}~\bibnamefont
  {Robertson}},\ }\href@noop {} {\bibfield  {journal} {\bibinfo  {journal}
  {Nano Lett.}\ }\textbf {\bibinfo {volume} {6}},\ \bibinfo {pages}
  {1107--1112} (\bibinfo {year} {2006})}\BibitemShut {NoStop}%
\bibitem [{\citenamefont {Jinkins}\ \emph
  {et~al.}(2019{\natexlab{b}})\citenamefont {Jinkins}, \citenamefont {Chan},
  \citenamefont {Jacobberger}, \citenamefont {Berson},\ and\ \citenamefont
  {Arnold}}]{Jinkins_2019}%
  \BibitemOpen
  \bibfield  {author} {\bibinfo {author} {\bibfnamefont {K.~R.}\ \bibnamefont
  {Jinkins}}, \bibinfo {author} {\bibfnamefont {J.}~\bibnamefont {Chan}},
  \bibinfo {author} {\bibfnamefont {R.~M.}\ \bibnamefont {Jacobberger}},
  \bibinfo {author} {\bibfnamefont {A.}~\bibnamefont {Berson}},\ and\ \bibinfo
  {author} {\bibfnamefont {M.~S.}\ \bibnamefont {Arnold}},\ }\href@noop {}
  {\bibfield  {journal} {\bibinfo  {journal} {Adv. Electron. Mater.}\ }\textbf
  {\bibinfo {volume} {5}},\ \bibinfo {pages} {1800593} (\bibinfo {year}
  {2019}{\natexlab{b}})}\BibitemShut {NoStop}%
\bibitem [{\citenamefont {Vaillancourt}\ \emph {et~al.}(2008)\citenamefont
  {Vaillancourt}, \citenamefont {Zhang}, \citenamefont {Vasinajindakaw},
  \citenamefont {Xia}, \citenamefont {Lu}, \citenamefont {Han}, \citenamefont
  {Janzen}, \citenamefont {Shih}, \citenamefont {Jones}, \citenamefont
  {Stroder}, \citenamefont {Chen}, \citenamefont {Subbaraman}, \citenamefont
  {Chen}, \citenamefont {Berger},\ and\ \citenamefont
  {Renn}}]{Vaillancourt_2008_CNT_transistor}%
  \BibitemOpen
  \bibfield  {author} {\bibinfo {author} {\bibfnamefont {J.}~\bibnamefont
  {Vaillancourt}}, \bibinfo {author} {\bibfnamefont {H.}~\bibnamefont {Zhang}},
  \bibinfo {author} {\bibfnamefont {P.}~\bibnamefont {Vasinajindakaw}},
  \bibinfo {author} {\bibfnamefont {H.}~\bibnamefont {Xia}}, \bibinfo {author}
  {\bibfnamefont {X.}~\bibnamefont {Lu}}, \bibinfo {author} {\bibfnamefont
  {X.}~\bibnamefont {Han}}, \bibinfo {author} {\bibfnamefont {D.~C.}\
  \bibnamefont {Janzen}}, \bibinfo {author} {\bibfnamefont {W.}~\bibnamefont
  {Shih}}, \bibinfo {author} {\bibfnamefont {C.~S.}\ \bibnamefont {Jones}},
  \bibinfo {author} {\bibfnamefont {M.}~\bibnamefont {Stroder}}, \bibinfo
  {author} {\bibfnamefont {M.~Y.}\ \bibnamefont {Chen}}, \bibinfo {author}
  {\bibfnamefont {H.}~\bibnamefont {Subbaraman}}, \bibinfo {author}
  {\bibfnamefont {R.~T.}\ \bibnamefont {Chen}}, \bibinfo {author}
  {\bibfnamefont {U.}~\bibnamefont {Berger}},\ and\ \bibinfo {author}
  {\bibfnamefont {M.}~\bibnamefont {Renn}},\ }\href
  {https://doi.org/10.1063/1.3043682} {\bibfield  {journal} {\bibinfo
  {journal} {Applied Physics Letters}\ }\textbf {\bibinfo {volume} {93}},\
  \bibinfo {pages} {243301} (\bibinfo {year} {2008})},\ \Eprint
  {https://arxiv.org/abs/https://doi.org/10.1063/1.3043682}
  {https://doi.org/10.1063/1.3043682} \BibitemShut {NoStop}%
\bibitem [{\citenamefont {Xiao}\ \emph {et~al.}(2003)\citenamefont {Xiao},
  \citenamefont {Liu}, \citenamefont {Hu}, \citenamefont {Yu}, \citenamefont
  {Wang},\ and\ \citenamefont {Zhu}}]{Xiao_2003_transistor}%
  \BibitemOpen
  \bibfield  {author} {\bibinfo {author} {\bibfnamefont {K.}~\bibnamefont
  {Xiao}}, \bibinfo {author} {\bibfnamefont {Y.}~\bibnamefont {Liu}}, \bibinfo
  {author} {\bibfnamefont {P.}~\bibnamefont {Hu}}, \bibinfo {author}
  {\bibfnamefont {G.}~\bibnamefont {Yu}}, \bibinfo {author} {\bibfnamefont
  {X.}~\bibnamefont {Wang}},\ and\ \bibinfo {author} {\bibfnamefont
  {D.}~\bibnamefont {Zhu}},\ }\href {https://doi.org/10.1063/1.1589181}
  {\bibfield  {journal} {\bibinfo  {journal} {Applied Physics Letters}\
  }\textbf {\bibinfo {volume} {83}},\ \bibinfo {pages} {150--152} (\bibinfo
  {year} {2003})},\ \Eprint
  {https://arxiv.org/abs/https://doi.org/10.1063/1.1589181}
  {https://doi.org/10.1063/1.1589181} \BibitemShut {NoStop}%
\bibitem [{\citenamefont {Grechko}\ \emph {et~al.}(2014)\citenamefont
  {Grechko}, \citenamefont {Ye}, \citenamefont {Mehlenbacher}, \citenamefont
  {McDonough}, \citenamefont {Wu}, \citenamefont {Jacobberger}, \citenamefont
  {Arnold},\ and\ \citenamefont {Zanni}}]{hexagonal_CNT_bundles_Grechko_2014}%
  \BibitemOpen
  \bibfield  {author} {\bibinfo {author} {\bibfnamefont {M.}~\bibnamefont
  {Grechko}}, \bibinfo {author} {\bibfnamefont {Y.}~\bibnamefont {Ye}},
  \bibinfo {author} {\bibfnamefont {R.~D.}\ \bibnamefont {Mehlenbacher}},
  \bibinfo {author} {\bibfnamefont {T.~J.}\ \bibnamefont {McDonough}}, \bibinfo
  {author} {\bibfnamefont {M.}~\bibnamefont {Wu}}, \bibinfo {author}
  {\bibfnamefont {R.~M.}\ \bibnamefont {Jacobberger}}, \bibinfo {author}
  {\bibfnamefont {M.~S.}\ \bibnamefont {Arnold}},\ and\ \bibinfo {author}
  {\bibfnamefont {M.~T.}\ \bibnamefont {Zanni}},\ }\href@noop {} {\bibfield
  {journal} {\bibinfo  {journal} {ACS Nano}\ }\textbf {\bibinfo {volume} {8}},\
  \bibinfo {pages} {5383--5394} (\bibinfo {year} {2014})}\BibitemShut {NoStop}%
\bibitem [{\citenamefont {Virtanen}\ \emph {et~al.}(2020)\citenamefont
  {Virtanen}, \citenamefont {Gommers}, \citenamefont {Oliphant}, \citenamefont
  {Haberland}, \citenamefont {Reddy}, \citenamefont {Cournapeau}, \citenamefont
  {Burovski}, \citenamefont {Peterson}, \citenamefont {Weckesser},
  \citenamefont {Bright}, \citenamefont {{van der Walt}}, \citenamefont
  {Brett}, \citenamefont {Wilson}, \citenamefont {Millman}, \citenamefont
  {Mayorov}, \citenamefont {Nelson}, \citenamefont {Jones}, \citenamefont
  {Kern}, \citenamefont {Larson}, \citenamefont {Carey}, \citenamefont {Polat},
  \citenamefont {Feng}, \citenamefont {Moore}, \citenamefont {{VanderPlas}},
  \citenamefont {Laxalde}, \citenamefont {Perktold}, \citenamefont {Cimrman},
  \citenamefont {Henriksen}, \citenamefont {Quintero}, \citenamefont {Harris},
  \citenamefont {Archibald}, \citenamefont {Ribeiro}, \citenamefont
  {Pedregosa}, \citenamefont {{van Mulbregt}},\ and\ \citenamefont {{SciPy 1.0
  Contributors}}}]{Scipy}%
  \BibitemOpen
  \bibfield  {author} {\bibinfo {author} {\bibfnamefont {P.}~\bibnamefont
  {Virtanen}}, \bibinfo {author} {\bibfnamefont {R.}~\bibnamefont {Gommers}},
  \bibinfo {author} {\bibfnamefont {T.~E.}\ \bibnamefont {Oliphant}}, \bibinfo
  {author} {\bibfnamefont {M.}~\bibnamefont {Haberland}}, \bibinfo {author}
  {\bibfnamefont {T.}~\bibnamefont {Reddy}}, \bibinfo {author} {\bibfnamefont
  {D.}~\bibnamefont {Cournapeau}}, \bibinfo {author} {\bibfnamefont
  {E.}~\bibnamefont {Burovski}}, \bibinfo {author} {\bibfnamefont
  {P.}~\bibnamefont {Peterson}}, \bibinfo {author} {\bibfnamefont
  {W.}~\bibnamefont {Weckesser}}, \bibinfo {author} {\bibfnamefont
  {J.}~\bibnamefont {Bright}}, \bibinfo {author} {\bibfnamefont {S.~J.}\
  \bibnamefont {{van der Walt}}}, \bibinfo {author} {\bibfnamefont
  {M.}~\bibnamefont {Brett}}, \bibinfo {author} {\bibfnamefont
  {J.}~\bibnamefont {Wilson}}, \bibinfo {author} {\bibfnamefont {K.~J.}\
  \bibnamefont {Millman}}, \bibinfo {author} {\bibfnamefont {N.}~\bibnamefont
  {Mayorov}}, \bibinfo {author} {\bibfnamefont {A.~R.~J.}\ \bibnamefont
  {Nelson}}, \bibinfo {author} {\bibfnamefont {E.}~\bibnamefont {Jones}},
  \bibinfo {author} {\bibfnamefont {R.}~\bibnamefont {Kern}}, \bibinfo {author}
  {\bibfnamefont {E.}~\bibnamefont {Larson}}, \bibinfo {author} {\bibfnamefont
  {C.~J.}\ \bibnamefont {Carey}}, \bibinfo {author} {\bibfnamefont
  {{\.I}.}~\bibnamefont {Polat}}, \bibinfo {author} {\bibfnamefont
  {Y.}~\bibnamefont {Feng}}, \bibinfo {author} {\bibfnamefont {E.~W.}\
  \bibnamefont {Moore}}, \bibinfo {author} {\bibfnamefont {J.}~\bibnamefont
  {{VanderPlas}}}, \bibinfo {author} {\bibfnamefont {D.}~\bibnamefont
  {Laxalde}}, \bibinfo {author} {\bibfnamefont {J.}~\bibnamefont {Perktold}},
  \bibinfo {author} {\bibfnamefont {R.}~\bibnamefont {Cimrman}}, \bibinfo
  {author} {\bibfnamefont {I.}~\bibnamefont {Henriksen}}, \bibinfo {author}
  {\bibfnamefont {E.~A.}\ \bibnamefont {Quintero}}, \bibinfo {author}
  {\bibfnamefont {C.~R.}\ \bibnamefont {Harris}}, \bibinfo {author}
  {\bibfnamefont {A.~M.}\ \bibnamefont {Archibald}}, \bibinfo {author}
  {\bibfnamefont {A.~H.}\ \bibnamefont {Ribeiro}}, \bibinfo {author}
  {\bibfnamefont {F.}~\bibnamefont {Pedregosa}}, \bibinfo {author}
  {\bibfnamefont {P.}~\bibnamefont {{van Mulbregt}}},\ and\ \bibinfo {author}
  {\bibnamefont {{SciPy 1.0 Contributors}}},\ }\bibfield  {title} {\enquote
  {\bibinfo {title} {{{SciPy} 1.0: Fundamental Algorithms for Scientific
  Computing in Python}},}\ }\href {https://doi.org/10.1038/s41592-019-0686-2}
  {\bibfield  {journal} {\bibinfo  {journal} {Nature Methods}\ }\textbf
  {\bibinfo {volume} {17}},\ \bibinfo {pages} {261--272} (\bibinfo {year}
  {2020})}\BibitemShut {NoStop}%
\bibitem [{\citenamefont {Jacoboni}\ and\ \citenamefont
  {Reggiani}(1983)}]{Jacoboni_1983_EMC_rev}%
  \BibitemOpen
  \bibfield  {author} {\bibinfo {author} {\bibfnamefont {C.}~\bibnamefont
  {Jacoboni}}\ and\ \bibinfo {author} {\bibfnamefont {L.}~\bibnamefont
  {Reggiani}},\ }\href {https://doi.org/10.1103/RevModPhys.55.645} {\bibfield
  {journal} {\bibinfo  {journal} {Rev. Mod. Phys.}\ }\textbf {\bibinfo {volume}
  {55}},\ \bibinfo {pages} {645--705} (\bibinfo {year} {1983})}\BibitemShut
  {NoStop}%
\bibitem [{\citenamefont {Fauquembergue}\ \emph {et~al.}(1979)\citenamefont
  {Fauquembergue}, \citenamefont {Zimmermann}, \citenamefont {Kaszynski},\ and\
  \citenamefont {Constant}}]{diffusion_tensor_1979}%
  \BibitemOpen
  \bibfield  {author} {\bibinfo {author} {\bibfnamefont {R.}~\bibnamefont
  {Fauquembergue}}, \bibinfo {author} {\bibfnamefont {J.}~\bibnamefont
  {Zimmermann}}, \bibinfo {author} {\bibfnamefont {A.}~\bibnamefont
  {Kaszynski}},\ and\ \bibinfo {author} {\bibfnamefont {E.}~\bibnamefont
  {Constant}},\ }\href@noop {} {\bibfield  {journal} {\bibinfo  {journal} {J.
  Appl. Phys.}\ }\textbf {\bibinfo {volume} {51}},\ \bibinfo {pages}
  {1065--1071} (\bibinfo {year} {1979})}\BibitemShut {NoStop}%
\bibitem [{\citenamefont {Rengel}\ and\ \citenamefont
  {Martin}(2013)}]{diffusion_tensor_Rengel_2013}%
  \BibitemOpen
  \bibfield  {author} {\bibinfo {author} {\bibfnamefont {R.}~\bibnamefont
  {Rengel}}\ and\ \bibinfo {author} {\bibfnamefont {M.~J.}\ \bibnamefont
  {Martin}},\ }\href@noop {} {\bibfield  {journal} {\bibinfo  {journal} {J.
  Appl. Phys.}\ }\textbf {\bibinfo {volume} {114}},\ \bibinfo {pages} {143702}
  (\bibinfo {year} {2013})}\BibitemShut {NoStop}%
\bibitem [{\citenamefont {Nish}\ \emph {et~al.}(2007)\citenamefont {Nish},
  \citenamefont {Hwang}, \citenamefont {Doig},\ and\ \citenamefont
  {Nicholas}}]{polymer_wrapping_Nish_2007}%
  \BibitemOpen
  \bibfield  {author} {\bibinfo {author} {\bibfnamefont {A.}~\bibnamefont
  {Nish}}, \bibinfo {author} {\bibfnamefont {J.~Y.}\ \bibnamefont {Hwang}},
  \bibinfo {author} {\bibfnamefont {J.}~\bibnamefont {Doig}},\ and\ \bibinfo
  {author} {\bibfnamefont {R.~J.}\ \bibnamefont {Nicholas}},\ }\href@noop {}
  {\bibfield  {journal} {\bibinfo  {journal} {Nat. Nanotechnol.}\ }\textbf
  {\bibinfo {volume} {2}},\ \bibinfo {pages} {640--646} (\bibinfo {year}
  {2007})}\BibitemShut {NoStop}%
\bibitem [{\citenamefont {Wang}\ \emph {et~al.}(2017)\citenamefont {Wang},
  \citenamefont {Shea}, \citenamefont {Flach}, \citenamefont {McDonough},
  \citenamefont {Way}, \citenamefont {Zanni},\ and\ \citenamefont
  {Arnold}}]{Wang_2017_QuenchingSites}%
  \BibitemOpen
  \bibfield  {author} {\bibinfo {author} {\bibfnamefont {J.}~\bibnamefont
  {Wang}}, \bibinfo {author} {\bibfnamefont {M.~J.}\ \bibnamefont {Shea}},
  \bibinfo {author} {\bibfnamefont {J.~T.}\ \bibnamefont {Flach}}, \bibinfo
  {author} {\bibfnamefont {T.~J.}\ \bibnamefont {McDonough}}, \bibinfo {author}
  {\bibfnamefont {A.~J.}\ \bibnamefont {Way}}, \bibinfo {author} {\bibfnamefont
  {M.~T.}\ \bibnamefont {Zanni}},\ and\ \bibinfo {author} {\bibfnamefont
  {M.~S.}\ \bibnamefont {Arnold}},\ }\bibfield  {title} {\enquote {\bibinfo
  {title} {Role of defects as exciton quenching sites in carbon nanotube
  photovoltaics},}\ }\href {https://doi.org/10.1021/acs.jpcc.7b01005}
  {\bibfield  {journal} {\bibinfo  {journal} {The Journal of Physical Chemistry
  C}\ }\textbf {\bibinfo {volume} {121}},\ \bibinfo {pages} {8310--8318}
  (\bibinfo {year} {2017})},\ \Eprint
  {https://arxiv.org/abs/https://doi.org/10.1021/acs.jpcc.7b01005}
  {https://doi.org/10.1021/acs.jpcc.7b01005} \BibitemShut {NoStop}%
\end{thebibliography}
%

\end{document}